\documentclass[letterpaper,twocolumn,10pt]{article}
\PassOptionsToPackage{table}{xcolor}
\PassOptionsToPackage{hyphens}{url}
\usepackage{usenix2026_SOUPS}
\microtypecontext{spacing=nonfrench}

\usepackage{tikz}
\usepackage{amsmath}
\usepackage{enumitem}
\setlist{noitemsep}
\usepackage{colortbl}
\usepackage{tabularx}
\usepackage{siunitx}
\usepackage{booktabs}
\usepackage{wasysym}
\usepackage{graphicx}
\usepackage{subcaption}
\usepackage{etoolbox}

\newcommand{\definevar}[2]{%
  \expandafter\newcommand\csname var#1var\endcsname{#2}%
}
\newcommand{\var}[1]{\ifcsname var#1var\endcsname%
        \csname var#1var\endcsname%
    \else%
        TODO%
        \GenericWarning{LaTeX Warning: \noexpand Number/Variable {#1} is undefined. on input line \the\inputlineno}%
    \fi%
}

\usepackage[style=american,threshold=2]{csquotes}

\definevar{change_video_min}{89.20}
\definevar{change_video_max}{1397.00}
\definevar{change_video_median}{271.50}
\definevar{change_video_mean}{286.35}

\definevar{reset_video_min}{39.70}
\definevar{reset_video_max}{443.80}
\definevar{reset_video_median}{111.60}
\definevar{reset_video_mean}{115.87}

\definevar{diffLeastRankAndAmountOfCompared}{291}
\definevar{rankOfLeastPopularWebsiteCoded}{402}

\definevar{wellknowntotal}{6}
\definevar{wellknownchange}{4}
\definevar{wellknownrecovery}{2}

\definevar{account_activation_True_df_change}{0}
\definevar{account_activation_False_df_change}{96}
\definevar{account_activation_True_df_reset}{0}
\definevar{account_activation_False_df_reset}{109}
\definevar{auto_logged_in_True_df_change}{\textcolor{gray}{n/a}}
\definevar{auto_logged_in_False_df_change}{\textcolor{gray}{n/a}}
\definevar{auto_logged_in_True_df_reset}{51}
\definevar{auto_logged_in_False_df_reset}{58}
\definevar{captcha_True_df_change}{5}
\definevar{captcha_False_df_change}{91}
\definevar{captcha_True_df_reset}{16}
\definevar{captcha_False_df_reset}{93}
\definevar{change_info_True_df_change}{62}
\definevar{change_info_False_df_change}{34}
\definevar{change_info_True_df_reset}{64}
\definevar{change_info_False_df_reset}{45}
\definevar{confirm_password_True_df_change}{75}
\definevar{confirm_password_False_df_change}{21}
\definevar{confirm_password_True_df_reset}{75}
\definevar{confirm_password_False_df_reset}{34}
\definevar{confirmation_required_True_df_change}{0}
\definevar{confirmation_required_False_df_change}{96}
\definevar{confirmation_required_True_df_reset}{0}
\definevar{confirmation_required_False_df_reset}{109}
\definevar{embedded in page_True_df_change}{35}
\definevar{embedded in page_False_df_change}{61}
\definevar{embedded in page_True_df_reset}{59}
\definevar{embedded in page_False_df_reset}{50}
\definevar{feedback_both_True_df_change}{36}
\definevar{feedback_both_False_df_change}{75}
\definevar{feedback_both_True_df_reset}{42}
\definevar{feedback_both_False_df_reset}{69}
\definevar{hard to detect_True_df_change}{9}
\definevar{hard to detect_False_df_change}{87}
\definevar{hard to detect_True_df_reset}{1}
\definevar{hard to detect_False_df_reset}{108}
\definevar{html_autocomplete_both_True_df_change}{9}
\definevar{html_autocomplete_both_False_df_change}{102}
\definevar{html_autocomplete_both_True_df_reset}{\textcolor{gray}{n/a}}
\definevar{html_autocomplete_both_False_df_reset}{\textcolor{gray}{n/a}}
\definevar{html_autocomplete_change_True_df_change}{15}
\definevar{html_autocomplete_change_False_df_change}{96}
\definevar{html_autocomplete_change_True_df_reset}{\textcolor{gray}{n/a}}
\definevar{html_autocomplete_change_False_df_reset}{\textcolor{gray}{n/a}}
\definevar{html_autocomplete_login_True_df_change}{43}
\definevar{html_autocomplete_login_False_df_change}{68}
\definevar{html_autocomplete_login_True_df_reset}{\textcolor{gray}{n/a}}
\definevar{html_autocomplete_login_False_df_reset}{\textcolor{gray}{n/a}}
\definevar{logout_all_devices_True_df_change}{8}
\definevar{logout_all_devices_False_df_change}{88}
\definevar{logout_all_devices_True_df_reset}{7}
\definevar{logout_all_devices_False_df_reset}{102}
\definevar{new_password_True_df_change}{95}
\definevar{new_password_False_df_change}{1}
\definevar{new_password_True_df_reset}{109}
\definevar{new_password_False_df_reset}{0}
\definevar{no_policy_no_strength_meter_no_password_generator_no_pw_leak_check_True_df_change}{111}
\definevar{no_policy_no_strength_meter_no_password_generator_no_pw_leak_check_False_df_change}{\textcolor{gray}{n/a}}
\definevar{no_policy_no_strength_meter_no_password_generator_no_pw_leak_check_True_df_reset}{111}
\definevar{no_policy_no_strength_meter_no_password_generator_no_pw_leak_check_False_df_reset}{\textcolor{gray}{n/a}}
\definevar{old/current_password_True_df_change}{80}
\definevar{old/current_password_False_df_change}{16}
\definevar{old/current_password_True_df_reset}{\textcolor{gray}{n/a}}
\definevar{old/current_password_False_df_reset}{\textcolor{gray}{n/a}}
\definevar{partly_logout_from_devices_True_df_change}{12}
\definevar{partly_logout_from_devices_False_df_change}{84}
\definevar{partly_logout_from_devices_True_df_reset}{4}
\definevar{partly_logout_from_devices_False_df_reset}{105}
\definevar{policy_displayed_True_df_change}{64}
\definevar{policy_displayed_False_df_change}{32}
\definevar{policy_displayed_True_df_reset}{81}
\definevar{policy_displayed_False_df_reset}{28}
\definevar{pop-up_True_df_change}{44}
\definevar{pop-up_False_df_change}{52}
\definevar{pop-up_True_df_reset}{16}
\definevar{pop-up_False_df_reset}{93}
\definevar{pre_verification_True_df_change}{11}
\definevar{pre_verification_False_df_change}{85}
\definevar{pre_verification_True_df_reset}{\textcolor{gray}{n/a}}
\definevar{pre_verification_False_df_reset}{\textcolor{gray}{n/a}}
\definevar{pre_verification_page_True_df_change}{9}
\definevar{pre_verification_page_False_df_change}{87}
\definevar{pre_verification_page_True_df_reset}{\textcolor{gray}{n/a}}
\definevar{pre_verification_page_False_df_reset}{\textcolor{gray}{n/a}}
\definevar{pw_generator_True_df_change}{2}
\definevar{pw_generator_False_df_change}{94}
\definevar{pw_generator_True_df_reset}{2}
\definevar{pw_generator_False_df_reset}{107}
\definevar{pw_leak/breach_check_True_df_change}{1}
\definevar{pw_leak/breach_check_False_df_change}{95}
\definevar{pw_leak/breach_check_True_df_reset}{1}
\definevar{pw_leak/breach_check_False_df_reset}{108}
\definevar{pw_reuse_prevention_True_df_change}{48}
\definevar{pw_reuse_prevention_False_df_change}{48}
\definevar{pw_reuse_prevention_True_df_reset}{47}
\definevar{pw_reuse_prevention_False_df_reset}{62}
\definevar{pwr<->pwc_True_df_change}{15}
\definevar{pwr<->pwc_False_df_change}{96}
\definevar{pwr<->pwc_True_df_reset}{2}
\definevar{pwr<->pwc_False_df_reset}{109}
\definevar{rba_capchta_True_df_change}{7}
\definevar{rba_capchta_False_df_change}{89}
\definevar{rba_capchta_True_df_reset}{\textcolor{gray}{n/a}}
\definevar{rba_capchta_False_df_reset}{\textcolor{gray}{n/a}}
\definevar{rba_email_True_df_change}{19}
\definevar{rba_email_False_df_change}{77}
\definevar{rba_email_True_df_reset}{\textcolor{gray}{n/a}}
\definevar{rba_email_False_df_reset}{\textcolor{gray}{n/a}}
\definevar{rba_not_provided_True_df_change}{68}
\definevar{rba_not_provided_False_df_change}{28}
\definevar{rba_not_provided_True_df_reset}{\textcolor{gray}{n/a}}
\definevar{rba_not_provided_False_df_reset}{\textcolor{gray}{n/a}}
\definevar{rba_sms_True_df_change}{2}
\definevar{rba_sms_False_df_change}{94}
\definevar{rba_sms_True_df_reset}{\textcolor{gray}{n/a}}
\definevar{rba_sms_False_df_reset}{\textcolor{gray}{n/a}}
\definevar{redirect_True_df_change}{44}
\definevar{redirect_False_df_change}{52}
\definevar{redirect_True_df_reset}{77}
\definevar{redirect_False_df_reset}{32}
\definevar{redirect_main_page_True_df_change}{4}
\definevar{redirect_main_page_False_df_change}{92}
\definevar{redirect_main_page_True_df_reset}{28}
\definevar{redirect_main_page_False_df_reset}{81}
\definevar{redirect_not_provided_True_df_change}{51}
\definevar{redirect_not_provided_False_df_change}{45}
\definevar{redirect_not_provided_True_df_reset}{32}
\definevar{redirect_not_provided_False_df_reset}{77}
\definevar{redirect_other_True_df_change}{26}
\definevar{redirect_other_False_df_change}{70}
\definevar{redirect_other_True_df_reset}{43}
\definevar{redirect_other_False_df_reset}{66}
\definevar{redirect_same_page_True_df_change}{15}
\definevar{redirect_same_page_False_df_change}{81}
\definevar{redirect_same_page_True_df_reset}{6}
\definevar{redirect_same_page_False_df_reset}{103}
\definevar{redirect_same_page_no_visual_feedback_True_df_change}{\textcolor{gray}{n/a}}
\definevar{redirect_same_page_no_visual_feedback_False_df_change}{111}
\definevar{redirect_same_page_no_visual_feedback_True_df_reset}{2}
\definevar{redirect_same_page_no_visual_feedback_False_df_reset}{109}
\definevar{redirect_session_expire_(auto_log_out)_True_df_change}{21}
\definevar{redirect_session_expire_(auto_log_out)_False_df_change}{75}
\definevar{redirect_session_expire_(auto_log_out)_True_df_reset}{\textcolor{gray}{n/a}}
\definevar{redirect_session_expire_(auto_log_out)_False_df_reset}{\textcolor{gray}{n/a}}
\definevar{secure_password_tips_True_df_change}{20}
\definevar{secure_password_tips_False_df_change}{76}
\definevar{secure_password_tips_True_df_reset}{19}
\definevar{secure_password_tips_False_df_reset}{90}
\definevar{short display time_True_df_change}{18}
\definevar{short display time_False_df_change}{78}
\definevar{short display time_True_df_reset}{8}
\definevar{short display time_False_df_reset}{101}
\definevar{show_pw_indicator_True_df_change}{53}
\definevar{show_pw_indicator_False_df_change}{43}
\definevar{show_pw_indicator_True_df_reset}{66}
\definevar{show_pw_indicator_False_df_reset}{43}
\definevar{signin_notification_True_df_change}{17}
\definevar{signin_notification_False_df_change}{79}
\definevar{signin_notification_True_df_reset}{6}
\definevar{signin_notification_False_df_reset}{103}
\definevar{strength_meter_True_df_change}{30}
\definevar{strength_meter_False_df_change}{66}
\definevar{strength_meter_True_df_reset}{27}
\definevar{strength_meter_False_df_reset}{82}
\definevar{visual_feedback_True_df_change}{78}
\definevar{visual_feedback_False_df_change}{18}
\definevar{visual_feedback_True_df_reset}{74}
\definevar{visual_feedback_False_df_reset}{35}
\definevar{account_activation_equality_count}{95}
\definevar{account_activation_equality_percent}{100}
\definevar{confirmation_required_equality_count}{95}
\definevar{confirmation_required_equality_percent}{100}
\definevar{no_policy_no_strength_meter_no_password_generator_no_pw_leak_check_equality_count}{95}
\definevar{no_policy_no_strength_meter_no_password_generator_no_pw_leak_check_equality_percent}{100}
\definevar{pw_generator_equality_count}{95}
\definevar{pw_generator_equality_percent}{100}
\definevar{pw_leak/breach_check_equality_count}{95}
\definevar{pw_leak/breach_check_equality_percent}{100}
\definevar{redirect_same_page_no_visual_feedback_equality_count}{94}
\definevar{redirect_same_page_no_visual_feedback_equality_percent}{99}
\definevar{new_password_equality_count}{94}
\definevar{new_password_equality_percent}{99}
\definevar{partly_logout_from_devices_equality_count}{87}
\definevar{partly_logout_from_devices_equality_percent}{92}
\definevar{hard to detect_equality_count}{86}
\definevar{hard to detect_equality_percent}{91}
\definevar{logout_all_devices_equality_count}{86}
\definevar{logout_all_devices_equality_percent}{91}
\definevar{strength_meter_equality_count}{85}
\definevar{strength_meter_equality_percent}{89}
\definevar{captcha_equality_count}{83}
\definevar{captcha_equality_percent}{87}
\definevar{change_info_equality_count}{81}
\definevar{change_info_equality_percent}{85}
\definevar{show_pw_indicator_equality_count}{81}
\definevar{show_pw_indicator_equality_percent}{85}
\definevar{secure_password_tips_equality_count}{80}
\definevar{secure_password_tips_equality_percent}{84}
\definevar{policy_displayed_equality_count}{80}
\definevar{policy_displayed_equality_percent}{84}
\definevar{redirect_same_page_equality_count}{80}
\definevar{redirect_same_page_equality_percent}{84}
\definevar{confirm_password_equality_count}{77}
\definevar{confirm_password_equality_percent}{81}
\definevar{signin_notification_equality_count}{76}
\definevar{signin_notification_equality_percent}{80}
\definevar{redirect_session_expire_(auto_log_out)_equality_count}{74}
\definevar{redirect_session_expire_(auto_log_out)_equality_percent}{78}
\definevar{short display time_equality_count}{72}
\definevar{short display time_equality_percent}{76}
\definevar{pw_reuse_prevention_equality_count}{72}
\definevar{pw_reuse_prevention_equality_percent}{76}
\definevar{redirect_main_page_equality_count}{70}
\definevar{redirect_main_page_equality_percent}{74}
\definevar{pop-up_equality_count}{60}
\definevar{pop-up_equality_percent}{63}
\definevar{redirect_other_equality_count}{60}
\definevar{redirect_other_equality_percent}{63}
\definevar{visual_feedback_equality_count}{60}
\definevar{visual_feedback_equality_percent}{63}
\definevar{embedded in page_equality_count}{55}
\definevar{embedded in page_equality_percent}{58}
\definevar{redirect_not_provided_equality_count}{54}
\definevar{redirect_not_provided_equality_percent}{57}
\definevar{redirect_equality_count}{53}
\definevar{redirect_equality_percent}{56}
\definevar{feedback_both_equality_count}{52}
\definevar{feedback_both_equality_percent}{55}
\definevar{auto_logged_in_equality_count}{51}
\definevar{auto_logged_in_equality_percent}{54}
\definevar{old/current_password_equality_count}{16}
\definevar{old/current_password_equality_percent}{17}
\definevar{total_URLS}{111}

\usepackage[acronym]{glossaries}
\newacronym{pwc}{PWC}{password change}
\newacronym{pwr}{PWR}{password reset}

\newbox{\orcidaffilbox}
\sbox{\orcidaffilbox}{\large\includegraphics[height=1.7ex]{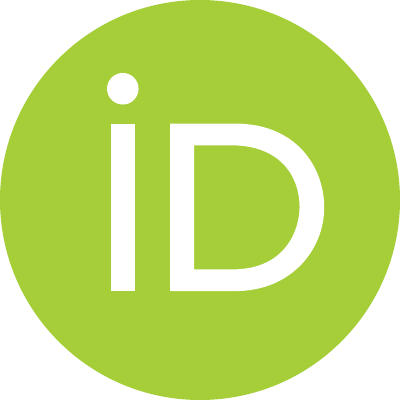}}
\newcommand{\orcidID}[1]{\href{https://orcid.org/#1}{\usebox{\orcidaffilbox}}}

\newcommand{\recitem}[1]{%
  \par\noindent
  \hangindent=1.6em
  \hangafter=1
  \hspace{0.5em}\makebox[1em][l]{$\bullet$}#1\par
}

\usepackage{paracol}
\globalcounter{figure}
\begin{document}

\date{}
\title{\Large \bf An Analysis of the Security, Usability, and Automation Capabilities of\\Password Update Processes on Top-Ranked Websites}

\def\plainauthor{Alexander Krause, Jacques Suray, Lea Schmüser, Marten Oltrogge, Oliver Wiese, Maximillian Golla, Sascha Fahl}

\author{
{\rm Alexander Krause\,\orcidID{0000-0003-2993-2568}, Jacques Suray\,\orcidID{0009-0007-8595-3706}, Lea Schmüser\,\orcidID{0009-0009-8991-4098},}
\and
{\rm Marten Oltrogge\,\orcidID{0000-0003-4920-308X}, Oliver Wiese\,\orcidID{0000-0003-2483-327X}, Maximillian Golla\,\orcidID{0000-0003-2204-2132}, and Sascha Fahl\,\orcidID{0000-0002-5644-3316}}\\
CISPA Helmholtz Center for Information Security
} %

\maketitle
\pagestyle{empty}
\thispagestyle{empty}
\begin{abstract}
Password updates are a critical part of the password lifecycle and are recommended following exposure of reused passwords or suspected compromise.
However, password update processes are often cumbersome, require manual password creation, and involve inconsistent website workflows that hinder reliable automation by password managers.

In this work, we conduct the first in-depth, systematic analysis of 111 password update processes deployed on top-ranked websites.
We provide novel insights into their overall security, usability, and automation capabilities, and contribute to authentication security research by improving the understanding of password update processes.
Websites often deploy highly diverse, complex, and confusing password update processes that are not supported by password managers.
Processes are often challenging to use, and end-users struggle to transfer experience and knowledge across websites.
Notably, security measures designed to enhance security often hinder password manager automation.
We conclude our work by discussing our findings and giving recommendations for web developers, the web standardization community, and security researchers.

\end{abstract}

\section{Introduction}\label{sec:intro}

User authentication security and usability have been an enormous challenge for the computer security community for decades. 
Most users have to manage tens or even hundreds of online accounts.
While authentication technologies are diverse and research and industry have suggested alternatives to password-based authentication, such as passkeys~\cite{lyastani-20-kingslayer,lassak-21-webauthn-misconceptions}, passwords remain a critical authentication mechanism:
Passwords are more inclusive and fail-safe; they can easily be shared with others, and end-users do not need additional devices, apps, or tokens to authenticate.

However, previous research~\cite{bonneau-12-the-quest, florencio-14-admins, thomas-17-leak} and a plethora of password breaches~\cite{franceschi-16-linkedin-surface-web, brewster-16-dropbox-bcrypt, thielman-16-yahoo-breach, hunt-13-haveibeenpwned} have illustrated that password-based authentication is full of challenges for end-users, service operators, and software developers.
Having to manage unique and hard-to-guess passwords for all of their accounts is unrealistic for end-users~\cite{florencio-14-finite-effort}, which is why they have developed various coping strategies like using easy-to-memorize (and guess) passwords~\cite{ur-15-added-at-the-end, pearman-17-habitat} or reusing similar passwords across accounts~\cite{nisenoff-23-reuse, golla-18-reuse-notification, wash-16-pw-choice}.
Although password managers can help end-users, previous research illustrates challenges concerning adoption~\cite{pearman-19-use-pw-manager, ray-21-pw-manager-old, mayer-22-pwm-edu}, usability~\cite{oesch-22-started-pw-manager, zibaei-22-random, kablo-24-audit-pwm-ea}, and security~\cite{lyastani-18-pw-manager, oesch-20-pw-managers, aonzo-18-phishing-pw-android}.
Developers struggle to store passwords securely~\cite{naiakshina-19-pw-storage-freelancer} and have trouble implementing password manager-friendly authentication flows~\cite{huaman-21-technical-issues, oesch-20-pw-managers}.
Service providers deploy password policies~\cite{lee-22-policy-top-website, al-roomi-23-login-policies} that can negatively impact overall account security~\cite{inglesant-10-cost-policies, habib-18-expiration}.

While password creation and management have been extensively studied~\cite{stobert-14-pw-life-cycle, ur-17-data-driven-pm, pearman-17-habitat, pearman-19-use-pw-manager, lee-22-policy-top-website, klivan-23-cred-management}, little attention has been paid to the password update processes implemented by websites.
End-users might need to update their passwords for multiple reasons: (1)~they may have forgotten the password and need to use a website's reset feature~\cite{bonneau-15-lies-and-account-recovery}; 
 (2)~they might need to go through an account remediation process~\cite{neil-21-remediation-advice, walsh-21-account-remediation}, for example, in response to a breach~\cite{nist-25-sp800-63b-4}; (3)~they might need to unshare the password~\cite{song-19-sharing-workplace, obada-obieh-20-account-sharing, alam-23-pw-sharing}, for example, after a breakup or in response to employee offboarding; (4)~they update it as a precaution, in response to a login notification email reporting about suspicious account activity~\cite{markert-24-login-notification}.
While the above list of reasons to update a password is not exhaustive, it highlights the need for secure and user-friendly password update processes to enhance account security.

Web password updates remain non-automated and inconsistent, forcing users to locate forms and navigate multi-step procedures that vary across sites. While password managers excel at generating and storing credentials~\cite{oesch-20-pw-managers, oesch-22-started-pw-manager}, automated password updates remain limited.
Dashlane's attempt at one-click updates illustrates this challenge: introduced (2014)~\cite{newton-14-dashlane-pwchanger}, revamped (2021)~\cite{peters-21-dashlane-pwchanger-upgrade}, then discontinued (2022)~\cite{dashlane-22-pwchanger-discontinued} due to maintenance complexity~\cite{sumner-23-dashlane-reasons}.
The W3C's password update standardization effort, accessible via a well-known URL~\cite{mondello-24-well-known}, could streamline access to update forms for users and tools.
However, our findings reveal a limited adoption of this proposed standard and persistently complex update processes.

To date, the literature lacks a systematic analysis of password update processes on the web.
We aim to contribute to a deeper understanding of the challenges posed by password update processes and give recommendations to enhance the security, usability, and automation capabilities.
In this work, we answer and discuss the following research questions:
\begin{enumerate}[leftmargin=*,label=RQ\arabic*,format=\bfseries,itemsep=0pt,partopsep=0pt,parsep=0pt,labelwidth=2pt]
    \item\label{RQ:state-of-the-art} [State~of~the~Art] %
    \emph{How do websites implement password update processes?}
    We aim to gain in-depth insights into password update processes and investigate their impact on end-users and password managers.
    \item\label{RQ:obstacles} [Challenges and Obstacles] \emph{What are common security, usability, and automation challenges and obstacles end-users and password managers face when interacting with password update processes?} 
    We aim to identify and understand the challenges to contribute towards more secure and usable processes in the future.
    \item\label{RQ:recommendations} [Recommendations] \emph{How can websites be better supported to deploy more secure, easy-to-use, and more easily automated password update processes?}
    We give recommendations for service providers, password managers, and the research community.
\end{enumerate}

\noindent \textbf{Contributions and Key Findings.}
We analyze password update processes on \var{total_URLS} top-ranked Tranco websites~\cite{lepochat-19-tranco}, focusing on security, usability, and automation capabilities.
We found that update forms are often buried deep in the account settings, requiring up to six clicks, and use inconsistent designs.
Some forms omit password confirmation fields; many lack visible password requirements, and some enforce conflicting policies during recovery versus regular password changes.
Most do not provide confirmations, and many fail to send notifications, enabling stealthy, unauthorized changes.
Automation faces barriers: Most forms lack ``autocomplete'' attributes~\cite{chromium-16-password-form}, few support the W3C draft~\cite{mondello-24-well-known}, and some deploy a CAPTCHA or RBA~\cite{wiefling-19-rba-in-the-wild}, blocking automated tools.
Overall, many update processes on top-ranked websites are complex, confusing, and inconsistent. 
They might negatively impact end-user authentication usability, hinder password manager support, and decrease authentication security.

\section{Reasons to Update Passwords}\label{sec:rel_reasons_for_changing_a_password}
We distinguish password changes, where users still know their current password, from password recovery, where they have lost access to it.\\

\noindent \textbf{Password Change (PWC).}\label{bp:rel_password_change}
Regular password changes were once a common piece of advice and part of many password policies in the past.
However, prior work has illustrated that making password changes mandatory is frustrating for users and results in weaker passwords~\cite{chiasson-15-pw-expiration, ncsc-16-password-expiry, habib-18-expiration}.
Yet there are many good reasons to change a password:
For example, a password change is recommended after a breach~\cite{franceschi-16-linkedin-surface-web} and often part of the account remediation process~\cite{neil-21-remediation-advice, walsh-21-account-remediation}.
Other reasons for changing a password include unsharing a password~\cite{song-19-sharing-workplace, obada-obieh-20-account-sharing, alam-23-pw-sharing}, improving account security hygiene~\cite{pullman-19-password-checkup}, and responding to suspicious login notifications~\cite{markert-24-login-notification}.\\

\noindent \textbf{Password Reset (PWR).}\label{bp:rel_password_reset}
In case users forget their password or lose access otherwise, most websites deploy password reset processes to allow users to set a new one.
In contrast to a password change, a password reset involves different means of (fallback) authentication to confirm the identity of the user~\cite{lassak-24-fallback-lt, reeder-11-secondary-auth}, for example, by clicking a link in an email, entering a code received via SMS, or answering personal knowledge questions~\cite{bonneau-15-lies-and-account-recovery, golla-15-ashley-pkq}.

\section{Related Work}\label{sec:related-work}
\label{sec:rel}
Prior work on passwords is extensive; we focus on two areas: breach alerting and account remediation~\cite{thomas-19-pw-checkup, huang-22-chrome-breach-notification, neil-21-remediation-advice, markert-24-login-notification}, e.g., by notifying users about their passwords' appearance in a password leak and how to respond and secure their accounts.
Our work builds upon this by investigating the steps commonly required to change or reset a password.
The second major area focuses on password managers and automatic password changes~\cite{oesch-20-pw-managers, newton-14-dashlane-pwchanger}, such as W3C's attempt~\cite{mondello-24-well-known} to standardize an API to facilitate automatic changes.
Our work builds on this by documenting obstacles and discussing solutions that could help automate the process.\\

\noindent \textbf{Breach Alerting \& Account Remediation.}\label{bp:rel_password_breach_alerting}
All major browsers~\cite{nguyen-18-firefox-monitor, thomas-19-pw-checkup, esposito-20-safari-compromised-pw}, and a variety of password managers~\cite{fey-18-1password-watchtower} notify users if their passwords appear in a data breach.
A popular free service is ``Have I Been Pwned?'' by Tony Hunt~\cite{hunt-13-haveibeenpwned}, where users can learn about their appearance in a data breach by entering their email address.
Academia has proposed protocols for compromised credential checking services~\cite{li-19-checking-credentials, pal-21-might-i-get-pwned, pasquini-24-leak-checking} and tried to improve security notifications~\cite{zou-19-useful-breach-notifications, huang-22-chrome-breach-notification}.
For example, Mayer et al.~\cite{mayer-21-a-bit-angry} studied individuals' awareness and perception of data breaches.
They found that 74\% of the reported breaches were new to the participants.
Golla et al.~\cite{golla-18-reuse-notification} studied password reuse notifications and noted the importance of ecosystem-level strategies, including password managers and convenient password update mechanisms.
Zou et al.~\cite{zou-19-might-be-affected} analyzed 161~breach notifications.
They found that notifications are often difficult to read, sometimes downplay the consequences, and frequently provide insufficient information on the urgency and prioritization of actions.
Regarding account remediation, Neil et al.~\cite{neil-21-remediation-advice} analyzed advice from 57~websites.
They found that highly ranked websites and those that had already experienced a data breach provided more comprehensive advice for their users.
Similarly, Walsh et al.~\cite{walsh-21-account-remediation} studied advice given by top websites from seven different countries.
They found that the provided instructions often miss essential steps and that users are forced to rely on third-party sources to resolve their problems.
Oh et al.~\cite{oh-25-checkup} evaluated factors influencing the ease of updating passwords, e.g., after a breach.
Beyond remediation advice, Daffalla et al.~\cite{daffalla-23-asi} examined account security interfaces~(ASIs), finding them inconsistent, insecure, and ineffective.
Bhattacharya et al.~\cite{bhattacharya-26-asi} further showed that ASI logs and notifications remain difficult to navigate, even for expert users.\\

\noindent \textbf{Password Managers \& Automatic Changes.}
\label{bp:rel_password_managers}
Prior work has studied password manager limitations: Lyastani et al.~\cite{lyastani-18-pw-manager} quantified security, Huaman et al.~\cite{huaman-21-technical-issues} documented HTML compatibility, and Oesch and Ruoti~\cite{oesch-20-pw-managers} revealed autocomplete inconsistencies and non-standard field handling.

There have been attempts to automate or increase the usability of changing a password online.
In 2014, Dashlane offered a one-click solution to automatically change passwords on popular websites~\cite{newton-14-dashlane-pwchanger}.
In 2021, they revised the tool to run on the client side~\cite{peters-21-dashlane-pwchanger-upgrade} to accommodate two-factor authentication~(2FA) and RBA prompts caused by their outdated server-side solution.
Interestingly, in 2022 Dashlane decided to discontinue the changer~\cite{dashlane-22-pwchanger-discontinued}, because ``[\dots] \textit{it was an extremely complex feature that was prone to breaking as sites changed their change password forms \& processes}''~\cite{sumner-23-dashlane-reasons}.

\textit{Well-known} Uniform Resource Identifiers (URIs) are defined by the IETF in RFC~8615~\cite{nottingham-19-rfc8615}.
It addresses the need for the consistent availability of services or information at specific URLs across servers.
In 2018, an engineer working for Apple proposed the concept of ``a well-known URL for changing passwords''~\cite{oconnor-18-well-known}, i.e., ``example.com/.well-known/change-password.'' %
The goal is to make the password updating process easier to use by automatically directing users of tools, such as password managers, to the password update form, thereby easing the hassle of navigating to it manually.
Shortly after, Apple Safari, Google Chrome, Microsoft Edge, and certain third-party password managers implemented the proposal.
In 2021, the W3C officially published a working draft~\cite{mondello-24-well-known} standardizing the concept.
Dashlane extended this concept with their own proposal, focusing on password managers~\cite{dashlane-21-well-known-proposal}. They propose a similar, well-known URL, but with an extension that supports a lightweight automatic password change protocol. This protocol should enable password managers to automatically change a password by providing the username, current password, and new password. They also introduce a minimal approach to handle two-factor authentication and CAPTCHAs.
Closely related to password changing APIs are password policy description languages~\cite{horsch-16-pw-markup,gautam-22-pw-policy-language} and password manager quirks~\cite{apple-20-manager-quirks}, which we do not consider here.

\section{Method}\label{sec:method}

\begin{table*}[!t]
\centering
\begingroup
\fontsize{10}{12}\selectfont
\setlength{\abovecaptionskip}{0pt}
\setlength{\belowcaptionskip}{3pt}
\setlength{\tabcolsep}{2pt}
\renewcommand{\arraystretch}{1.0}
\setlength{\extrarowheight}{0pt}
\providecommand{\subattr}[1]{\hspace{0.6em}#1}

\caption{Attributes and their descriptions in the \textbf{analysis protocol} for the password change and reset processes. Findings for the update processes, navigation path lengths, and the ``password change'' W3C draft adoption are presented in Section~\ref{sec:results}.}
\label{tab:analysis_protocoll}
\smallskip

\rowcolors{4}{white}{gray!10}
\begin{tabularx}{\textwidth}{@{}>{\raggedleft\arraybackslash}p{1.35em} >{\raggedright\arraybackslash}p{0.195\textwidth} >{\raggedright\arraybackslash}X >{\centering\arraybackslash}p{2.8em} >{\centering\arraybackslash}p{2.8em}@{}}
\toprule
\textbf{ID} & \textbf{Attribute} & \textbf{Description} & \textbf{\gls{pwc}} & \textbf{\gls{pwr}} \\
\midrule

\multicolumn{5}{@{}c@{}}{\textbf{Initialization}} \\
\textbf{1} & RBA~Email & An additional risk-based email authentication was performed. & $\newmoon$ & $\fullmoon$ \\
\textbf{2} & RBA~CAPTCHA & An additional risk-based CAPTCHA authentication was performed. & $\newmoon$ & $\fullmoon$ \\
\textbf{3} & RBA~SMS & An additional risk-based SMS authentication was performed. & $\newmoon$ & $\fullmoon$ \\
\textbf{4} & Autocomplete Login & Is True if the attribute is set to `email' or `username' for the user field and `current-password' for the password field. & $\newmoon$ & $\fullmoon$ \\

\multicolumn{5}{@{}c@{}}{\textbf{Password Update}} \\
\textbf{5} & Opposite Process & The \gls{pwc} process references the \gls{pwr} process and vice versa. & $\newmoon$ & $\newmoon$ \\
\textbf{6} & Pre-Verification & To enter the password update form, the user must verify their identity again. & $\newmoon$ & $\fullmoon$ \\
\textbf{7} & Current PW Field & The current password is required to update the password. & $\newmoon$ & $\fullmoon$ \\
\textbf{8} & Confirm PW Field & To update the password, the new password must be confirmed once again. & $\newmoon$ & $\newmoon$ \\
\textbf{9} & Autocomplete \gls{pwc} & Is True if the attribute is set to `current-password' for the current-password field and to `new-password' for the `new' and `confirm' password field. & $\newmoon$ & $\fullmoon$ \\
\textbf{10} & PW Policy & Website provides information about its own password composition policy. & $\newmoon$ & $\newmoon$ \\
\textbf{11} & PW Visibility Toggle & Users have the option to toggle the visibility of the passwords they enter. & $\newmoon$ & $\newmoon$ \\
\textbf{12} & PW Generator & The website has a built-in function to generate a random password. & $\newmoon$ & $\newmoon$ \\
\textbf{13} & PW Strength Meter & Website provides feedback about the password strength. & $\newmoon$ & $\newmoon$ \\
\textbf{14} & PW Breach Check & Website checks whether user data appears in known data breaches. & $\newmoon$ & $\newmoon$ \\
\textbf{15} & Secure PW Tips & Website provides advice for strong passwords. & $\newmoon$ & $\newmoon$ \\
\textbf{16} & CAPTCHA & At least one CAPTCHA had to be solved before or after the password update. & $\newmoon$ & $\newmoon$ \\
\textbf{17} & PW Reuse Prevention & Updating the password to a previously used password is actively prevented. & $\newmoon$ & $\newmoon$ \\
\textbf{18} & Logout All Devices & Website communicates that all sessions expire when updating the password. & $\newmoon$ & $\newmoon$ \\
\textbf{19} & Logout Partially & Website communicates that some sessions expire, excluding the user's session. & $\newmoon$ & $\newmoon$ \\
\textbf{20} & Expire Current Session & Without any prior information, the user is logged out after the update. & $\newmoon$ & $\fullmoon$ \\
\textbf{21} & Automatic Login & The user is automatically logged in after updating their password. & $\fullmoon$ & $\newmoon$ \\

\multicolumn{5}{@{}c@{}}{\textbf{Feedback}} \\
\textbf{22} & Visual Feedback & The website visually communicates a successful password update. & $\newmoon$ & $\newmoon$ \\
\textbf{23} & \subattr{Pop-up} & The feedback is displayed in a temporary or closable pop-up. & $\newmoon$ & $\newmoon$ \\
\textbf{24} & \subattr{Short-Display-Time} & The feedback is only visible for a very short amount of time. & $\newmoon$ & $\newmoon$ \\
\textbf{25} & \subattr{Embedded in Page} & The feedback is persistent and embedded in the webpage. & $\newmoon$ & $\newmoon$ \\
\textbf{26} & \subattr{Hard to Detect} & The feedback is hard to notice due to the size, arrangement on the page, or color. & $\newmoon$ & $\newmoon$ \\
\textbf{27} & Redirect & There is an automatic redirect after updating the password. & $\newmoon$ & $\newmoon$ \\
\textbf{28} & \subattr{Redirect Main Page} & The redirect leads to the main page. & $\newmoon$ & $\newmoon$ \\
\textbf{29} & \subattr{Redirect Same Page} & The redirect directs to the same page without changing the URL. & $\newmoon$ & $\newmoon$ \\
\textbf{30} & \subattr{Redirect Other} & The redirect leads neither to the main page nor to the same page. & $\newmoon$ & $\newmoon$ \\
\textbf{31} & Sign-in Notification & A sign-in notification is sent to the user's email address. & $\newmoon$ & $\newmoon$ \\
\textbf{32} & PW Update Notification & A password update notification is sent to the user's email address. & $\newmoon$ & $\newmoon$ \\
\bottomrule
\multicolumn{5}{@{}>{\centering\arraybackslash}p{.975\textwidth}@{}}{$\newmoon$ Attribute analyzed.\quad $\fullmoon$ Attribute not applicable and therefore not part of the analysis.} \\
\end{tabularx}
\endgroup
\end{table*}

Next, we describe our methodology for analyzing password change and reset implementations using new accounts. Finally, we discuss study limitations and ethics. Our dataset and protocol are available in our replication package~\cite{krause-26-pw-change-artifact}.

\subsection{Data Collection}
\label{sec:method_data_collection}

We sampled websites from the Tranco~list~\cite{lepochat-19-tranco} and analyzed a subset of the top-ranked websites.\footnote{List: \url{https://tranco-list.eu/list/X57KN}, as of March 1, 2023.}
We began by analyzing websites that allowed us to create an account and did not meet our exclusion criteria (outlined below).
In total, we created accounts on \var{total_URLS} websites between September~2024 and February~2025.
A complete list of the websites can be found in Table~\ref{tab:analyzed-websites} in Appendix~\ref{appendix:full-websites-list}.
For websites that required SMS-based verification in addition to an email address, we used a US-based or German phone number to create the account when necessary.
To avoid any IP-based geo-blocking, we utilized a commercial VPN to spoof our location.\\

\textbf{Account Creation Process.}
\label{bp:method_account_creation_process}
Before we began creating accounts, we prepared all the necessary information, including first and last name, birthdate, address, phone number, email address, and generic answers to security questions. 
If the website had a password policy, we generated a random password that complied with it.
If the website had no policy, we generated a 20-character random password.
During account creation, we did not enroll in non-mandatory 2FA.\\

\textbf{Exclusion Criteria.}
\label{bp:method_exclusion_criteria}
We excluded websites for which we could not create a free account, such as those that require a subscription.
Furthermore, we had to exclude websites where our account was banned, for example, because it was not a unique or real person.
In addition, we excluded all websites that were not available in English or that we could not translate to English using Google Chrome's internal website translation feature~\cite{google-26-translate}.
In line with previous work~\cite{amft-23-recover-mfa}, we also excluded websites that would have required us to sign a contract or upload a passport.
In total, we had to exclude \var{diffLeastRankAndAmountOfCompared} websites (between the highest Tranco rank~1 and the lowest rank of~\var{rankOfLeastPopularWebsiteCoded} that we decided to consider in our analysis).\\

\noindent \textbf{Protocol.}
\label{bp:method_analysis_protocol}
We developed our initial analysis protocol based on related work~\cite{huaman-21-technical-issues, amft-23-recover-mfa, buettner-23-rba-fallback, wiefling-19-rba-in-the-wild} and discussions among the team while jointly analyzing \gls{pwc} and \gls{pwr} processes.
Our protocol focuses on attributes that help us evaluate the user experience, automation capabilities, and security implications.
After evaluating the first 10 websites, we refined and expanded the protocol to include 32 attributes.
To minimize data-collection errors and facilitate reproducibility, we used a custom-built form.
We recorded our screens while analyzing the \gls{pwc} and \gls{pwr} processes on the websites.
Our implementation of the analysis protocol is available in the replication package.
The final protocol checks multiple attributes depending on whether it is a \gls{pwc} or \gls{pwr} process (see Table~\ref{tab:analysis_protocoll}).
We started with the login form, as a login is often required to change a password.
The form also typically allows the start of a \gls{pwr} process.
We analyzed automation capabilities, e.g., various HTML attributes and security measures deployed by the service, e.g., RBA, or the presence of a password strength meter.
We also evaluated website feedback and notifications for successful password updates, such as those sent by email or SMS.
In addition, we analyzed whether and how websites implement the well-known URL for changing passwords W3C draft by appending \texttt{/.well-known/change-password} to the domains.\\

\noindent \textbf{Password Update Processes.}
\label{bp:method_coding}
Two researchers independently applied the protocol to each website using fresh Google Chrome profiles.
We first analyzed the \gls{pwc} process, following these steps: (1)~logging in; (2)~searching for the account settings to change the password; and (3)~changing the password following the service's instructions. 
We then looked at the \gls{pwr} process, following these steps: (1)~searching the website for the \gls{pwr} entry point; (2)~following the reset instructions; and (3)~setting a new password.
We also tested whether the website accepted the previous password or required the creation of a new one.
Screen recordings were collected during the analysis to support potential conflict resolution.
In the event of a conflict, the researchers jointly reviewed the recording and reached an agreement.
In total, we resolved 48 conflicts (1.6\% of all analyses) for the password change process and 197 conflicts (6.4\% of all analyses) for the password reset process.
We followed an iterative coding approach until no new password update process emerged in the final 10 websites~\cite{birks-22-grounded-theory,urquhart-13-grounded-theory}. 
As we resolved conflicts immediately when they emerged, we decided not to report an inter-coder agreement~\cite{wermke-22-oss-trust, mcdonald-19-irr, wermke-23-supply-chain, huaman-22-virtual, gutfleisch-22-non-usable-software, krause-23-pushed-by-accident}.
Multiple coders independently evaluated the workflows, achieving high initial agreement: PWC: 98.2\%, PWR: 94.7\%, combined: 96.3\%. We treated disagreements as conflicts and resolved them by reviewing the captured data. All disagreements resulted from human error rather than interpretive differences.

\subsection{Data Analysis}
\label{sec:method_data_analysis}
Multiple researchers jointly applied affinity mapping~\cite{beyer-97-contextual-design, pernice-18-affinity-diagramming}, a method for clustering information, to identify the steps needed to perform the \gls{pwc} and \gls{pwr}, and to organize the observations into potential automation, usability, and security issues.
Usability-related systematic analysis findings are grounded in established knowledge and practices~\cite{molich-90-dialogue, nielsen-90-heuristic, nielsen-94-usability-heuristic, nielsen-95-usability-inspection}.\\

\noindent \textbf{Quantitative vs.\ Qualitative.}
\label{bp:method_counts_vs_quanifiers}
We report the quantitative findings of our systematic analysis in Table~\ref{tab:obstacles:sec_measures}.
In addition, we use quantifiers, illustrated in Figure~\ref{fig:quantifier-figure} in Appendix~\ref{app:quantifiers}, to determine which qualitative findings are most relevant throughout the results (see Section~\ref{sec:results}) and discussion (see Section~\ref{sec:discussion}) sections.\\

\noindent \textbf{Limitations.}
Several limitations affect our work.
First, our sample does not cover all websites with user accounts; lower-ranked and less popular websites are underrepresented. We may therefore have missed insights outside our study scope, but followed established web and security measurement practices~\cite{amft-23-recover-mfa, demir-22-reproducibility, klein-22-exploits}.
Thus, we are confident that our dataset provides valuable insights to the community.
Second, if possible, we created accounts on the US versions of the websites (enforced by our browser language and IP address) and used US- and German-based phone numbers to receive SMS.
The behavior of large sites is known to change depending on the location, e.g., due to the GDPR.
However, we believe the effect on account management is marginal, particularly in relation to the properties relevant to our study. 
Our sample is also biased by the Tranco list (Eurocentrism) and may miss popular websites from other regions around the world, such as Asia, South America, or Africa.
Finally, websites are not static; developers regularly refine and improve their usability and security.
Our results are a snapshot of their current state.
Similarly, RBA and geo-restriction limit the reproducibility.
Although we recorded our sessions, reproducing the study may yield different findings for specific websites.
However, our general observations are reproducible.

\subsection{Ethics Statement}
\label{sec:ethics}
Our institution's ethical review board approved this work.
In our evaluation, we manually examined non-personal data. 
We took care not to cause unusual resource burdens on the websites and do not report any brands or website names with problematic processes in this work.
No help desk staff were involved in our experiments.
The involved researchers did not use their private accounts.
At no point was the researchers' personally identifiable information, particularly of the involved students, part of this research.
Reporting potential security vulnerabilities to the affected stakeholders is considered good scientific practice.
We examined our raw coding data for vulnerabilities.

\section{Results}\label{sec:results}

We now present the main findings of our systematic analysis.
This section is structured along (\textbf{\ref{RQ:state-of-the-art}}: State of the~Art and \textbf{\ref{RQ:obstacles}}:~Obstacles).
We provide details for the \gls{pwc} and \gls{pwr} processes with a particular focus on user experience, security, and the challenges faced by end-users and password managers.
To increase the readability of our findings reporting, we use quantifiers when referencing a number of websites illustrated in Figure~\ref{fig:quantifier-figure} in Appendix~\ref{app:quantifiers}.
We also use the IDs from Table~\ref{tab:obstacles:sec_measures} throughout the sections to relate to the analyzed and reported attributes.
Throughout, we report counts as $n$~($p\%$); unless noted otherwise, percentages are relative to the 96~websites offering a \gls{pwc} and the 109~offering a \gls{pwr}.

\subsection{Password Change and Reset Processes}
\label{sec:res_processes}

\begin{figure*}[!ht]
    \centering
    \includegraphics[width=1.0\textwidth]{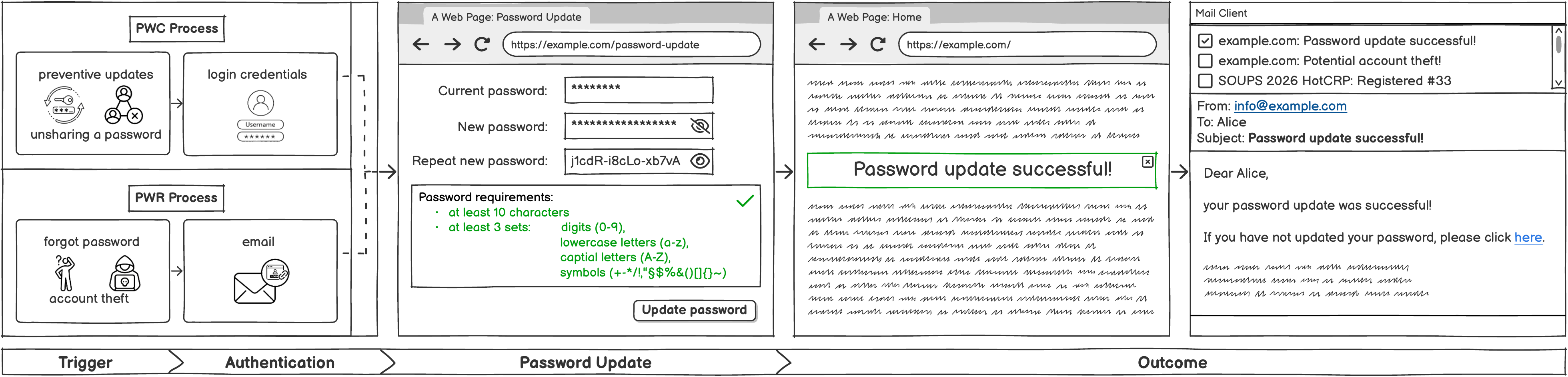}
    \vspace{-0.5cm}
    \caption{Illustration of a well-implemented password update (1) using \gls{pwc} and \gls{pwr} process \textbf{triggers}, (2) trigger-specific \textbf{authentication}, (3) \textbf{password update} forms including security features, and (4) the \textbf{outcome} communicated to the user.}
    \label{fig:process-overview}
\end{figure*}

\noindent \textbf{General Process.}
To sign in, all services required an identifier, such as an email address, username, phone number, or account ID, along with a password and, in some cases, a second factor.
Some websites deployed CAPTCHAs or RBA for additional security.
Overall, websites employed various login processes.

The \gls{pwc} forms on \var{old/current_password_True_df_change} of 96 websites (ID7) required us to enter the current password before we could set a new one or confirm our new password.
After updating the password, most websites provided visual confirmation.
The majority also sent notification emails or SMS; about half of the websites in the \gls{pwc} process and the majority in the \gls{pwr} processes sent HTTP redirects (\gls{pwc}: 44, 46\%; \gls{pwr}: 77, 71\%; ID27).
Some enforced immediate logouts by expiring the session during the \gls{pwc} process (\gls{pwc}: 21, 22\%; ID20).
Some services did not send any notifications after the password was changed.
The \gls{pwc} process was often similar to the \gls{pwr} process, and some websites even forwarded us to the \gls{pwr} process (\gls{pwc}: 15, 14\%; ID5).
Instead of providing a password update form, the website sent an email with instructions for resetting the password.
In contrast, only a few websites (\gls{pwr}: 2, 2\%; ID5) forwarded us from the \gls{pwr} to the \gls{pwc} process.
Three websites did not offer a \gls{pwc} process.
On four websites, we could not find a \gls{pwr} process.
Instead, they sent a link to a valid session in an email and redirected us to the \gls{pwc} form.
To start the \gls{pwr} process, websites often requested our email address.
Some requested our phone number, username, or account ID.
However, we also observed complex processes that require additional information, such as multiple identifiers (e.g., email and phone number).
While most \gls{pwr} processes were straightforward, we also observed implementations that required us to solve CAPTCHAs and answer questions. 
The email confirmation for \gls{pwr} varied a lot.
We sometimes received a link that created a session and redirected us to the \gls{pwc} form without being required to log in first. 
Other services asked us to copy or type a code or use a one-time password.
After we reached the \gls{pwr} form, the processes were often comparable to the \gls{pwc} processes in terms of security measures, visual feedback, and notifications.
Figure~\ref{fig:process-overview} illustrates a generic well-implemented process.

\subsection{User Experience and Interfaces} %
\label{sec:res_user_exp}
Below, we illustrate the user experiences of logging in, navigating to the password update form, and interacting with the form.
Those three aspects substantially impact the likelihood of successfully updating a password.\\

\noindent \textbf{Login \& Initialization (ID1--3,5).}
We could sign in on many websites by entering our email address and password.
No further interaction was required.
All but two of the websites that offered a \gls{pwr} process directly provided a ``Forgot Password'' link on their login form.
On some websites, we triggered RBA~\cite{wiefling-19-rba-in-the-wild} (email: 19, 20\%, ID1; CAPTCHAs: 7, 7\%, ID2; SMS: 2, 2\%, ID3) or other means of credential stuffing or bot protection~\cite{hunt-17-credential-stuffing}, which required additional user interactions, e.g., entering a code or solving a CAPTCHA. 
As is inherent with RBA, we encountered it in a non-deterministic way.
In such cases, we often had to provide a verification code received via email, SMS, or both.
Most websites allowed us to copy and paste the code.
However, we also had to manually type the code on some websites.
This was necessary because the website would sometimes drop parts of the code or put the complete code into a single-digit text field, which impacted the user experience.
Several websites implemented CAPTCHAs, including Google's reCAPTCHA~v3, which required users to complete tasks like image selection and text transcription.
Audio-based variants challenged users to identify clips containing birdsong from multiple samples.
A notable implementation involved dice-based challenges in which users matched target numbers to images of five dice. The correct selection required calculating the sum of visible dice faces, with multiple iterations per session. These complex implementations risked user frustration due to their time-consuming nature.
While CAPTCHAs occasionally appeared during password changes or resets (\gls{pwc}: 5, 5\%; \gls{pwr}: 16, 15\%; ID16), security questions remained uncommon.\\
\begin{figure}[!h]
    \centering
    \includegraphics[width=1\columnwidth]{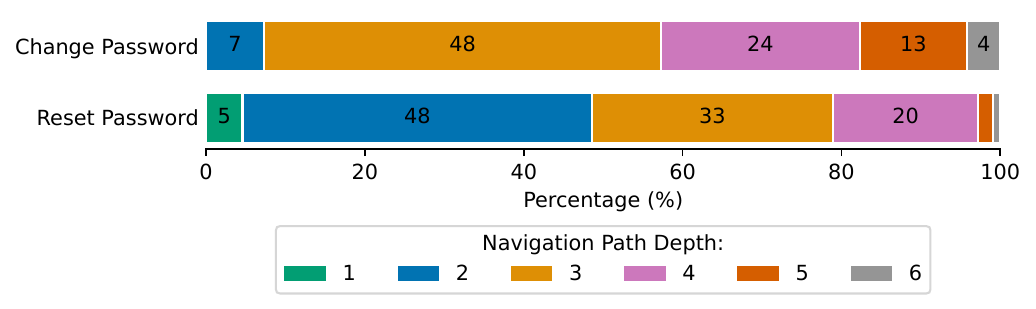}
    \caption{Distribution of the \textbf{navigation path depth} of \gls{pwc} and \gls{pwr} processes (higher depth often requires more effort). Example: \textit{Profile} $\rightarrow$ \textit{Advanced} $\rightarrow$ \textit{Security} $\rightarrow$ \textit{Password} (4).}
    \label{fig:pathDepth}
\end{figure}

\begin{figure*}[!ht]
     \centering
     \begin{subfigure}[b]{0.5\columnwidth}
         \centering
         \includegraphics[width=\columnwidth]{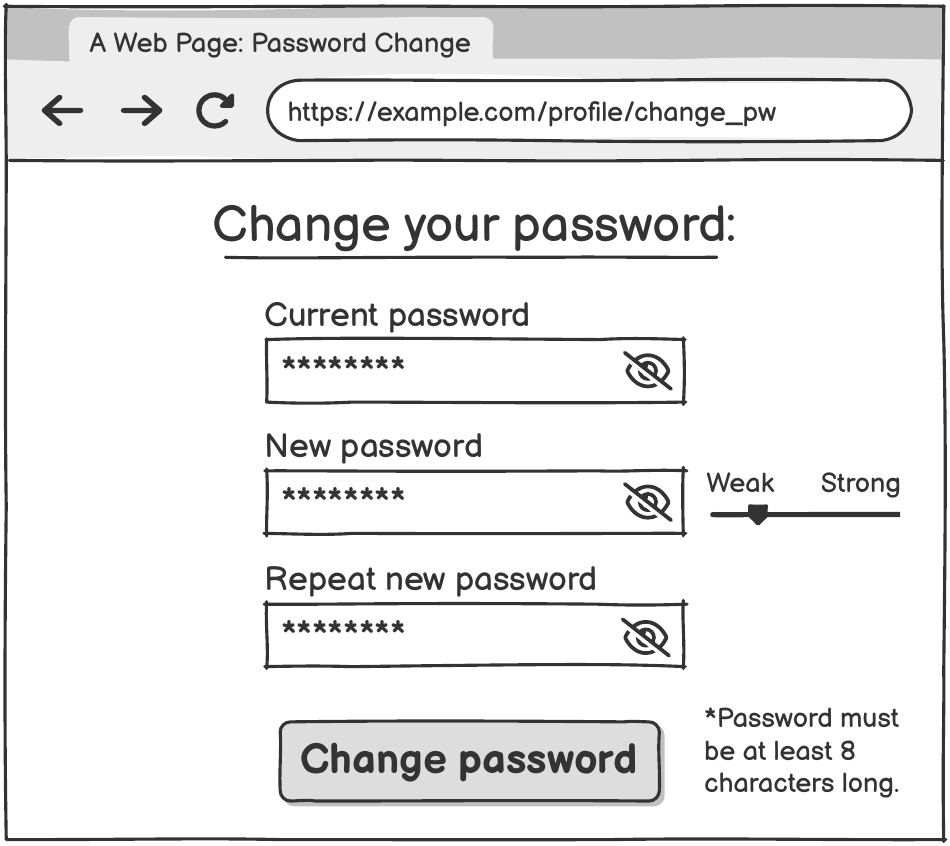}
         \caption{Dedicated Page with URL}
         \label{fig:form1}
     \end{subfigure}
     \begin{subfigure}[b]{0.5\columnwidth}
         \centering
         \includegraphics[width=\columnwidth]{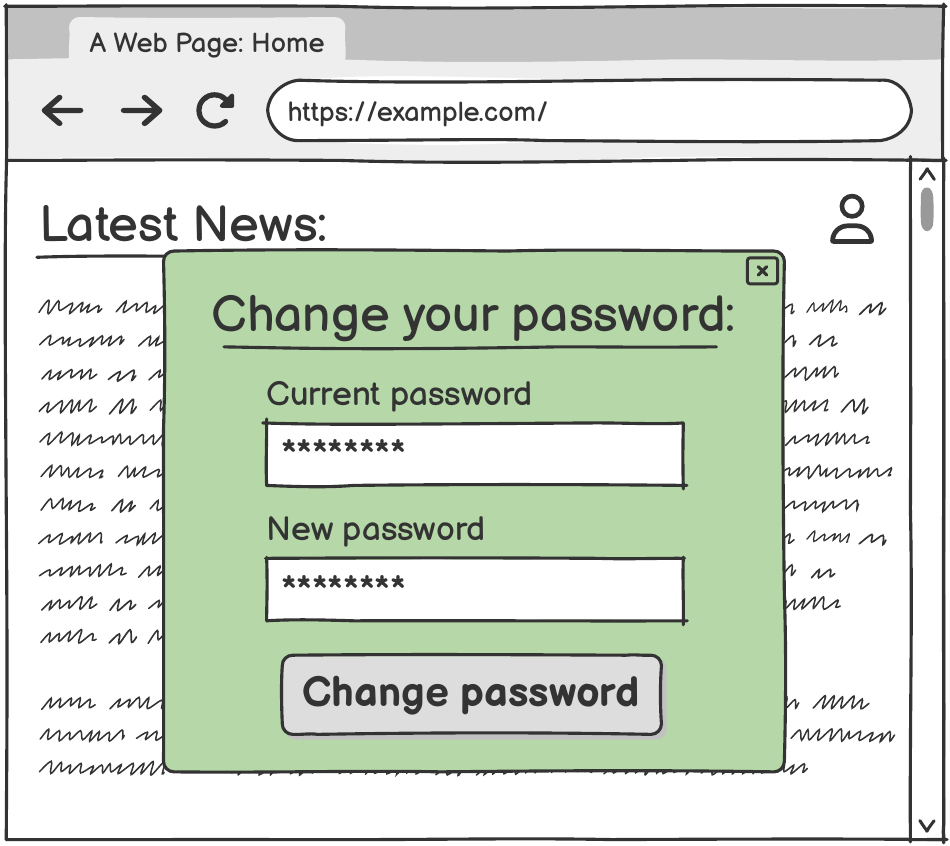}
         \caption{Modal without URL}
         \label{fig:form3}
     \end{subfigure}
          \begin{subfigure}[b]{0.5\columnwidth}
         \centering
         \includegraphics[width=\columnwidth]{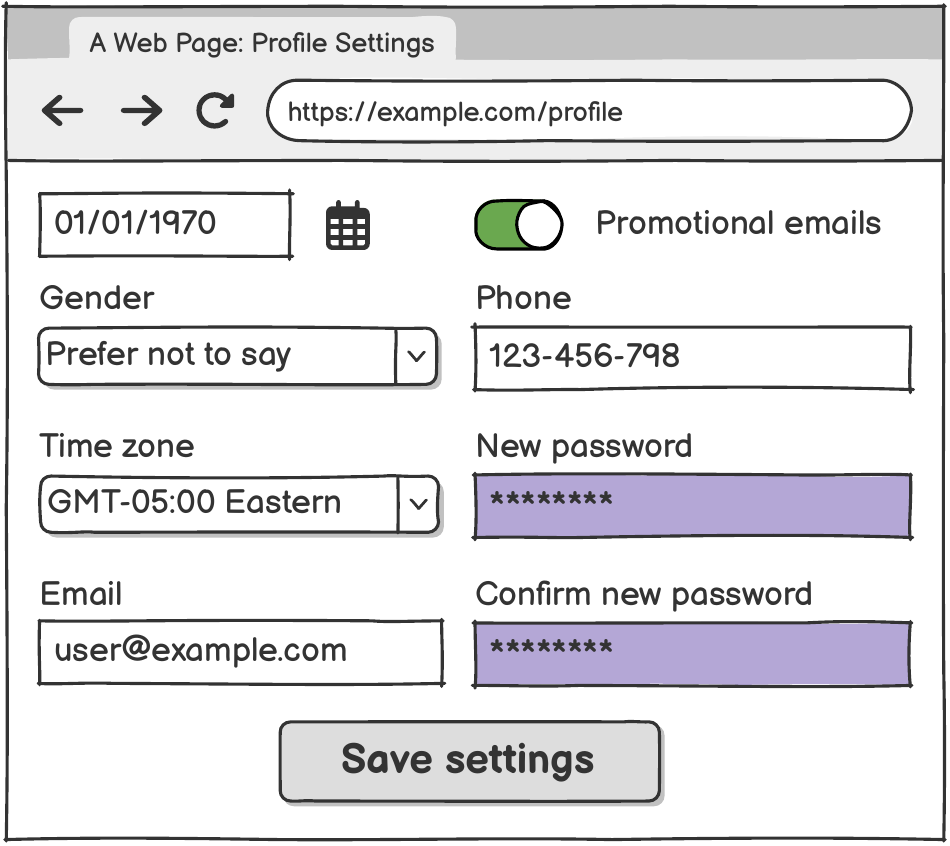}
         \caption{Profile Settings}
         \label{fig:form2}
     \end{subfigure}
          \begin{subfigure}[b]{0.5\columnwidth}
         \centering
         \includegraphics[width=\columnwidth]{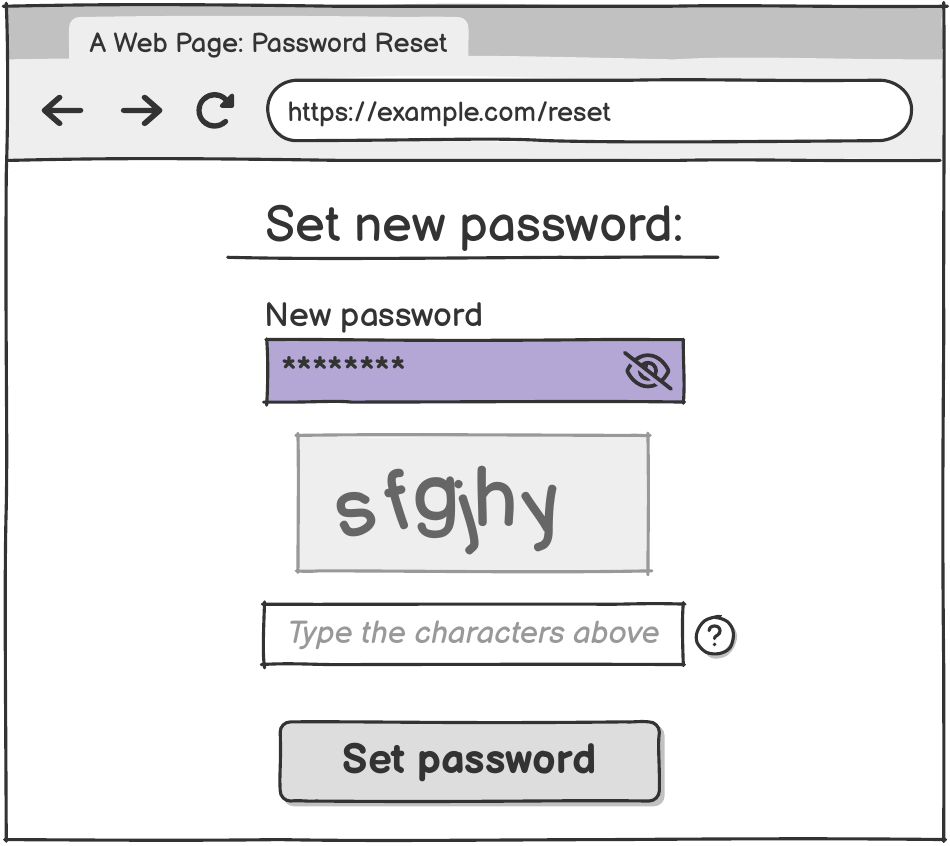}
         \caption{Reset with CAPTCHA}
         \label{fig:form4}
     \end{subfigure}
        \caption{Common \textbf{password forms} we observed in \gls{pwc} and \gls{pwr} processes. (a) illustrates a password change page accessible via a \textbf{dedicated URL}, (b) shows a \textbf{modal popup} of a web application that does not offer a URL, (c) displays \textbf{profile settings} that include some password fields, (d) shows a reset form with a \textbf{CAPTCHA} but no password confirmation field.}
        \label{fig:forms}
\end{figure*}

\noindent \textbf{Navigate to the Form.}
After logging in, we had to locate the \gls{pwc} form, which was generally challenging since websites often implemented confusing site paths.
Figure~\ref{fig:pathDepth} visualizes and compares the path depth for both processes.
In the \gls{pwc} process, the navigation path consisted of two to six steps, with three steps being the most frequent.
Despite similar labels, the name for the password update process differed from site to site.
For example, we frequently encountered: ``Authentication,'' ``Account Security,'' or ``Edit Profile.''
While three websites offered multiple paths to reach their password update forms, we had to use a search engine to find the starting point for at least one of them.
To trigger the \gls{pwr} process, we had to perform between one and six navigation steps, but mostly only two.
Typically, we had to access the login form and could trigger the \gls{pwr} by clicking on ``Forgot Password?'' or a similar label.
This disparity uniquely affects password updates, as resets are almost always initiated by a direct link provided in breach or recovery emails.
Conversely, \gls{pwc} workflows often lack standardization and consistent terminology, as evidenced by labels such as `Account Security' or `Change Password,' which can lead to usability issues.\\

\noindent \textbf{Interaction on the Form (ID7--8).}
We found different implementations of update forms in both the \gls{pwc} and \gls{pwr} processes.
A few are illustrated in Figure~\ref{fig:forms}.
Some were on a dedicated page, reachable via a dedicated URL.
Others were implemented as popups, either as a modal dialog within the page or as a separate window; i.e., the form cannot be reached directly via a URL. 
Most password change forms require users to enter their current password before setting a new one.
For both password change and password reset, most websites also require users to confirm the new password to avoid typos (\gls{pwc}: 75, 78\%; \gls{pwr}: 75, 69\%; ID8).
Still, confirmation was not required in all cases; some forms accepted the new password without asking users to retype it.
One particular website asked us to enter the current password in a second step after confirming the password change, which caused some frustration.
Since we used a password manager for our analysis, we had already updated the password manager entry for this website, leading to difficulties accessing the old entry in the vault, and thus, we could not complete the \gls{pwc} in the end.
Consequently, we had to reset the password using the \gls{pwr} process to regain access.

\subsection{Support Measures for Improved Security}
\label{sec:ressults_supportive_measures}
Table~\ref{tab:obstacles:sec_measures} shows the security features, which we detail below.\\

\begin{table}[!ht]
\centering
\begingroup
\fontsize{10}{12}\selectfont
\setlength{\abovecaptionskip}{0pt}
\setlength{\belowcaptionskip}{3pt}
\setlength{\tabcolsep}{2pt}
\renewcommand{\arraystretch}{1.0}
\setlength{\extrarowheight}{0pt}
\providecommand{\subattr}[1]{\hspace{0.6em}#1}

\caption{\textbf{Quantitative findings}. Because we reference the opposing processes (ID5), the total number of websites examined for each process is lower. The term \textit{equality} refers to the ratio of attributes implemented equally for both processes on the same site. Not every attribute is always applicable~(n/a).}
\label{tab:obstacles:sec_measures}
\smallskip

\rowcolors{4}{white}{gray!10}
\resizebox{\columnwidth}{!}{ \begin{tabular}{@{}>{\raggedleft\arraybackslash}p{1.35em} l S[table-format=3.0] S[table-format=3.0] S[table-format=3.0] S[table-format=3.0] >{\centering\arraybackslash}p{4.2em}@{}}
\toprule
\textbf{ID} & \textbf{Attribute} & \multicolumn{2}{c}{\textbf{Change (N=\var{pwr<->pwc_False_df_change})}} & \multicolumn{2}{c}{\textbf{Reset (N=\var{pwr<->pwc_False_df_reset})}} & \textbf{Equality} \\
\cmidrule(lr){3-4} \cmidrule(lr){5-6}
& & \multicolumn{1}{c}{\textbf{True}} & \multicolumn{1}{c}{\textbf{False}} & \multicolumn{1}{c}{\textbf{True}} & \multicolumn{1}{c}{\textbf{False}} & \textbf{\%}\\
\midrule

\multicolumn{7}{@{}c@{}}{\textbf{Initialization}} \\
\textbf{1} & RBA~Email & \var{rba_email_True_df_change} & \var{rba_email_False_df_change} & \var{rba_email_True_df_reset} & \var{rba_email_False_df_reset} & \textcolor{gray}{n/a}\\
\textbf{2} & RBA~CAPTCHA & \var{rba_capchta_True_df_change} & \var{rba_capchta_False_df_change} & \var{rba_capchta_True_df_reset} & \var{rba_capchta_False_df_reset} & \textcolor{gray}{n/a}\\
\textbf{3} & RBA~SMS & \var{rba_sms_True_df_change} & \var{rba_sms_False_df_change} & \var{rba_sms_True_df_reset} & \var{rba_sms_False_df_reset} & \textcolor{gray}{n/a}\\
\textbf{4} & Autocomplete Login & \var{html_autocomplete_login_True_df_change} & \var{html_autocomplete_login_False_df_change} & \var{html_autocomplete_login_True_df_reset} & \var{html_autocomplete_login_False_df_reset} & \textcolor{gray}{n/a}\\

\multicolumn{7}{@{}c@{}}{\textbf{Password Update}} \\
\textbf{5} & Opposite Process & \var{pwr<->pwc_True_df_change} & \var{pwr<->pwc_False_df_change} & \var{pwr<->pwc_True_df_reset} & \var{pwr<->pwc_False_df_reset} & \textcolor{gray}{n/a}\\
\textbf{6} & Pre-Verification & \var{pre_verification_True_df_change} & \var{pre_verification_False_df_change} & \var{pre_verification_True_df_reset} & \var{pre_verification_False_df_reset} & \textcolor{gray}{n/a}\\
\textbf{7} & Current PW Field & \var{old/current_password_True_df_change} & \var{old/current_password_False_df_change} & \var{old/current_password_True_df_reset} & \var{old/current_password_False_df_reset} & \textcolor{gray}{n/a}\\
\textbf{8} & Confirm PW Field & \var{confirm_password_True_df_change} & \var{confirm_password_False_df_change} & \var{confirm_password_True_df_reset} & \var{confirm_password_False_df_reset} & \var{confirmation_required_equality_percent} \\
\textbf{9} & Autocomplete \gls{pwc} & \var{html_autocomplete_change_True_df_change} & \var{html_autocomplete_change_False_df_change} & \var{html_autocomplete_change_True_df_reset} & \var{html_autocomplete_change_False_df_reset} & \textcolor{gray}{n/a}\\
\textbf{10} & PW Policy & \var{policy_displayed_True_df_change} & \var{policy_displayed_False_df_change} & \var{policy_displayed_True_df_reset} & \var{policy_displayed_False_df_reset} & \var{policy_displayed_equality_percent}\\
\textbf{11} & PW Visibility Toggle & \var{show_pw_indicator_True_df_change} & \var{show_pw_indicator_False_df_change} & \var{show_pw_indicator_True_df_reset} & \var{show_pw_indicator_False_df_reset} & \var{show_pw_indicator_equality_percent}\\
\textbf{12} & PW Generator & \var{pw_generator_True_df_change} & \var{pw_generator_False_df_change} & \var{pw_generator_True_df_reset} & \var{pw_generator_False_df_reset} & \var{pw_generator_equality_percent}\\
\textbf{13} & PW Strength Meter & \var{strength_meter_True_df_change} & \var{strength_meter_False_df_change} & \var{strength_meter_True_df_reset} & \var{strength_meter_False_df_reset} & \var{strength_meter_equality_percent}\\
\textbf{14} & PW Breach Check & \var{pw_leak/breach_check_True_df_change} & \var{pw_leak/breach_check_False_df_change} & \var{pw_leak/breach_check_True_df_reset} & \var{pw_leak/breach_check_False_df_reset} & \var{pw_leak/breach_check_equality_percent}\\
\textbf{15} & Secure PW Tips & \var{secure_password_tips_True_df_change} & \var{secure_password_tips_False_df_change} & \var{secure_password_tips_True_df_reset} & \var{secure_password_tips_False_df_reset} & \var{secure_password_tips_equality_percent}\\
\textbf{16} & CAPTCHA & \var{captcha_True_df_change} & \var{captcha_False_df_change} & \var{captcha_True_df_reset} & \var{captcha_False_df_reset} & \var{captcha_equality_percent}\\
\textbf{17} & PW Reuse Prevention & \var{pw_reuse_prevention_True_df_change} & \var{pw_reuse_prevention_False_df_change} & \var{pw_reuse_prevention_True_df_reset} & \var{pw_reuse_prevention_False_df_reset} & \var{pw_reuse_prevention_equality_percent}\\
\textbf{18} & Logout All Devices & \var{logout_all_devices_True_df_change} & \var{logout_all_devices_False_df_change} & \var{logout_all_devices_True_df_reset} & \var{logout_all_devices_False_df_reset} & \var{logout_all_devices_equality_percent}\\
\textbf{19} & Logout Partially & \var{partly_logout_from_devices_True_df_change} & \var{partly_logout_from_devices_False_df_change} & \var{partly_logout_from_devices_True_df_reset} & \var{partly_logout_from_devices_False_df_reset} & \var{partly_logout_from_devices_equality_percent} \\
\textbf{20} & Expire Current Session & \var{redirect_session_expire_(auto_log_out)_True_df_change} & \var{redirect_session_expire_(auto_log_out)_False_df_change} & \var{redirect_session_expire_(auto_log_out)_True_df_reset} & \var{redirect_session_expire_(auto_log_out)_False_df_reset} & \textcolor{gray}{n/a}\\
\textbf{21} & Automatic Login & \var{auto_logged_in_True_df_change} & \var{auto_logged_in_False_df_change} & \var{auto_logged_in_True_df_reset} & \var{auto_logged_in_False_df_reset} & \textcolor{gray}{n/a}\\

\multicolumn{7}{@{}c@{}}{\textbf{Feedback}} \\
\textbf{22} & Visual Feedback & \var{visual_feedback_True_df_change} & \var{visual_feedback_False_df_change} & \var{visual_feedback_True_df_reset} & \var{visual_feedback_False_df_reset} & \var{visual_feedback_equality_percent} \\
\textbf{23} & \subattr{Pop-up} & \var{pop-up_True_df_change} & \var{pop-up_False_df_change} & \var{pop-up_True_df_reset} & \var{pop-up_False_df_reset} & \var{pop-up_equality_percent}\\
\textbf{24} & \subattr{Short-Display-Time} & \var{short display time_True_df_change} & \var{short display time_False_df_change} & \var{short display time_True_df_reset} & \var{short display time_False_df_reset} & \var{short display time_equality_percent}\\
\textbf{25} & \subattr{Embedded in Page} & \var{embedded in page_True_df_change} & \var{embedded in page_False_df_change} & \var{embedded in page_True_df_reset} & \var{embedded in page_False_df_reset} & \var{embedded in page_equality_percent}\\
\textbf{26} & \subattr{Hard to Detect} & \var{hard to detect_True_df_change} & \var{hard to detect_False_df_change} & \var{hard to detect_True_df_reset} & \var{hard to detect_False_df_reset} & \var{hard to detect_equality_percent}\\
\textbf{27} & Redirect & \var{redirect_True_df_change} & \var{redirect_False_df_change} & \var{redirect_True_df_reset} & \var{redirect_False_df_reset} & \var{redirect_equality_percent}\\
\textbf{28} & \subattr{Redirect Main Page} & \var{redirect_main_page_True_df_change} & \var{redirect_main_page_False_df_change} & \var{redirect_main_page_True_df_reset} & \var{redirect_main_page_False_df_reset} & \var{redirect_main_page_equality_percent}\\
\textbf{29} & \subattr{Redirect Same Page} & \var{redirect_same_page_True_df_change} & \var{redirect_same_page_False_df_change} & \var{redirect_same_page_True_df_reset} & \var{redirect_same_page_False_df_reset} & \var{redirect_same_page_equality_percent}\\
\textbf{30} & \subattr{Redirect Other} & \var{redirect_other_True_df_change} & \var{redirect_other_False_df_change} & \var{redirect_other_True_df_reset} & \var{redirect_other_False_df_reset} & \var{redirect_other_equality_percent} \\
\textbf{31} & Sign-in Notification & \var{signin_notification_True_df_change} & \var{signin_notification_False_df_change} & \var{signin_notification_True_df_reset} & \var{signin_notification_False_df_reset} & \var{signin_notification_equality_percent} \\
\textbf{32} & PW Update Notification & \var{change_info_True_df_change} & \var{change_info_False_df_change} & \var{change_info_True_df_reset} & \var{change_info_False_df_reset} & \var{change_info_equality_percent}\\
\bottomrule
\end{tabular}}
\endgroup
\end{table}

\noindent \textbf{Session and Password Expiration (ID18--21).}
Only a few websites automatically logged us out of all devices after completing the \gls{pwc} process, leaving attackers with persistent session tokens (\gls{pwc}: 8, 8\%; ID18).
By comparison, session termination was equally rare but less critical during \gls{pwr}, where the reset invalidates previously used credentials (\gls{pwr}: 7, 6\%; ID18).
We could choose to terminate all active sessions, but more often we were informed of an automatic session expiration, or learned about it by navigating to a different page on the website after the \gls{pwc}, only to find ourselves logged out.
In one case, we had to confirm that all active sessions would become invalid because of the \gls{pwc}.
We also observed that on about half of the websites, the user was automatically logged in during the \gls{pwr} process (\gls{pwr}: 51, 47\%; ID21).  
Surprisingly, some websites implemented (the outdated and no longer recommended) automatic password expiration. 
On some sites, we could manually opt in to password expiration (see Figure~\ref{fig:expiration} in Appendix~\ref{app:website_examples}).
On a few, we were informed, e.g., ``\emph{the password will expire in 80~days.}''\\

\noindent \textbf{Password Policies (ID10).}
The majority of websites implement some password composition policy (\gls{pwc}: 64, 67\%; \gls{pwr}: 81, 74\%; ID10). 
Still, the implementation and policy varied across websites and, within the same website, between processes.
We discovered that password policies can vary based on whether the user is in the \gls{pwc} or \gls{pwr} process. Specifically, a password containing a hyphen ("-") was permitted in the \gls{pwr} process but was not allowed in the \gls{pwc} process on the same website.
We also observed websites that lacked a policy. %
Some displayed the policy requirements permanently, while others showed them after starting to type in the new password field.
The latter implementation often had the issue that the policy was not displayed when copying and pasting or auto-filling a password, e.g., generated from a password manager, into the text field.
Hence, websites rejected passwords that did not meet their composition requirements. 
We also found websites where the policy was displayed after clicking an additional help button, or when the entered password did not comply with it, i.e., we could not read the requirements before entering a password.
In total, most websites displayed the password policy during the \gls{pwr} process, whereas only a majority did so during the \gls{pwc} process.
Additionally, some policies prohibited using special characters like ``\%'', or ``\#''. 
More often, deployed password policies forced us to use a password of a certain length.
Particularly, very long passwords ($\geq$ 120 characters) were often rejected.
Overall, inconsistent implementations can degrade the user experience.
Enforcing unnecessarily stringent requirements can even break compatibility with passwords generated by password managers~\cite{huaman-21-technical-issues, oesch-20-pw-managers}, resulting in even more user frustration.
Despite the intention to ensure users are provided with strong passwords, we observed policies that mandate only a minimum of six characters, which is not in line with recommendations from academia~\cite{tan-20-pw-recommendations} and requirements published by organizations such as NIST~\cite{nist-25-sp800-63b-4}.\\

\noindent \textbf{Password Visibility Toggles (ID11).}
A toggle for password visibility allows users to review their password.
It is a crucial user interface component, as it can help to prevent typos and copy~\&~pasting errors, especially on mobile devices~\cite{melicher-16-mobile-passwords}.
The majority of websites implemented this feature for both processes (\gls{pwc}: 53, 55\%; \gls{pwr}: 66, 61\%; ID11).
We also found an unfavorable combination of a new password field without a confirmation field and without a visibility toggle.\\

\noindent \textbf{Password Leak \& Breach Checks (ID14).}
Only one website (\gls{pwc}: 1, 1\%; \gls{pwr}: 1, 1\%; ID14) effectively implemented a password leak check, i.e., a check for whether the user-generated password is part of a list of known breached credentials.
Two websites rejected passwords that were part of a known leak. One website disabled the password update button, while another required users to reset their passwords.
In contrast, another website routinely processed the update, resulting in a redirect to the main page, though with a warning that the password was not set.\\

\noindent \textbf{Password Generators (ID12).} A few websites (\gls{pwc}: 2, 2\%; \gls{pwr}: 2, 2\%; ID12) offered us to generate a secure and policy-compliant password, e.g., 
\enquote{wMwY\$f\allowbreak%
n\allowbreak%
S\allowbreak%
N\allowbreak%
H\allowbreak%
n\allowbreak%
B\allowbreak%
W\allowbreak%
t\allowbreak%
Iq\#wY\allowbreak%
iHmuA}
or \enquote{6chU\allowbreak%
mazy-9kro\allowbreak%
vnoe-lyum\allowbreak%
inal8}.
We observed differing views on secure passwords across websites, particularly regarding length and character set.
However, we did not observe any particular recommendation optimized for memorability or ease of typing and use.\\

\noindent \textbf{Password Strength Meters (ID13).}
While we found some websites (\gls{pwc}: 30, 31\%; \gls{pwr}: 27, 25\%; ID13) using a password strength meter to rate the guessability of a password during the \gls{pwc} and \gls{pwr} process, the used implementations differed in terms of rating and feedback.
We found websites using different visual representations, e.g., a single-colored three-step bar or a multi-colored five-step bar. In addition, the ranking was performed by calculating some ``entropy'' and was often based on character classes and length~(LUDS)~\cite{golla-18-psm}.\\

\begin{figure*}[!htbp]
     \centering
     \begin{subfigure}[b]{0.50\columnwidth}
         \centering
         \includegraphics[width=\columnwidth]{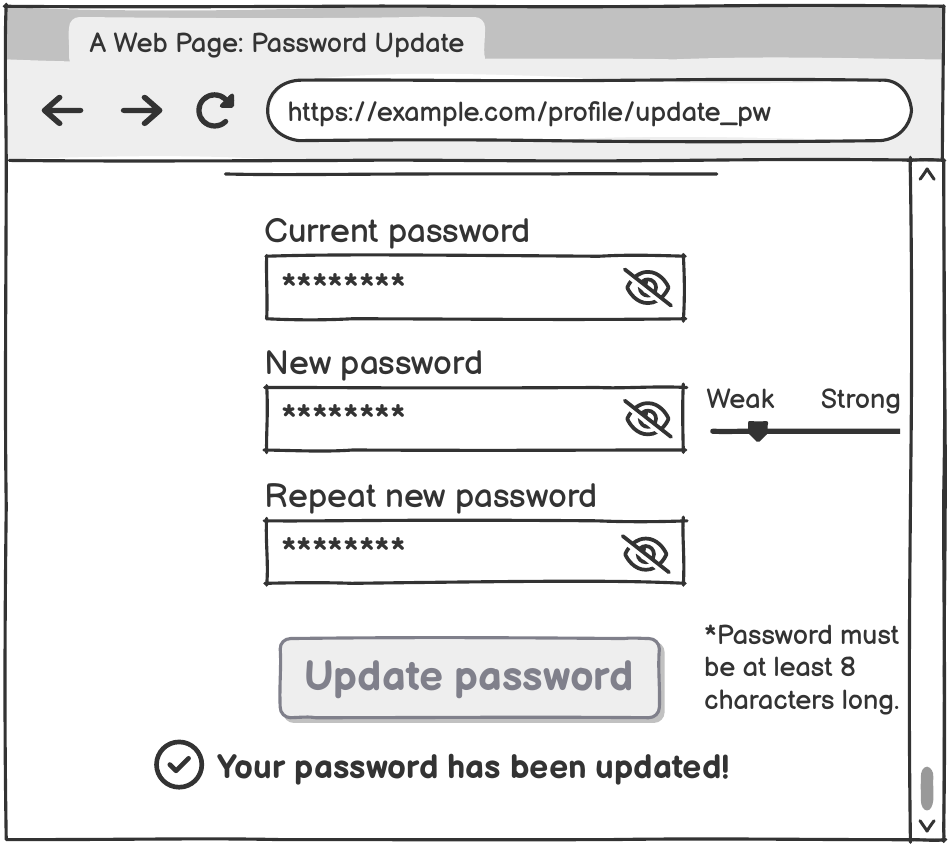}
         \caption{Explicit (ID25)}
         \label{fig:notification1}
     \end{subfigure}
     \begin{subfigure}[b]{0.50\columnwidth}
         \centering
         \includegraphics[width=\columnwidth]{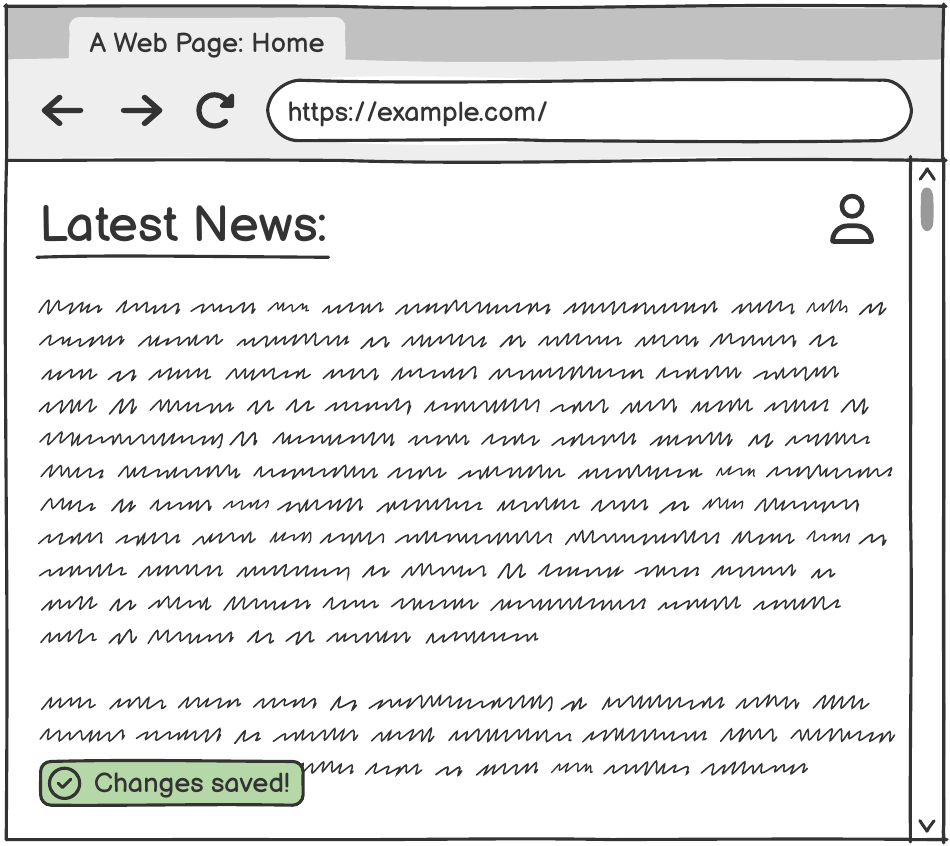}
         \caption{Tiny and Disappearing (ID23)}
         \label{fig:notification2}
     \end{subfigure}
          \begin{subfigure}[b]{0.50\columnwidth}
         \centering
         \includegraphics[width=\columnwidth]{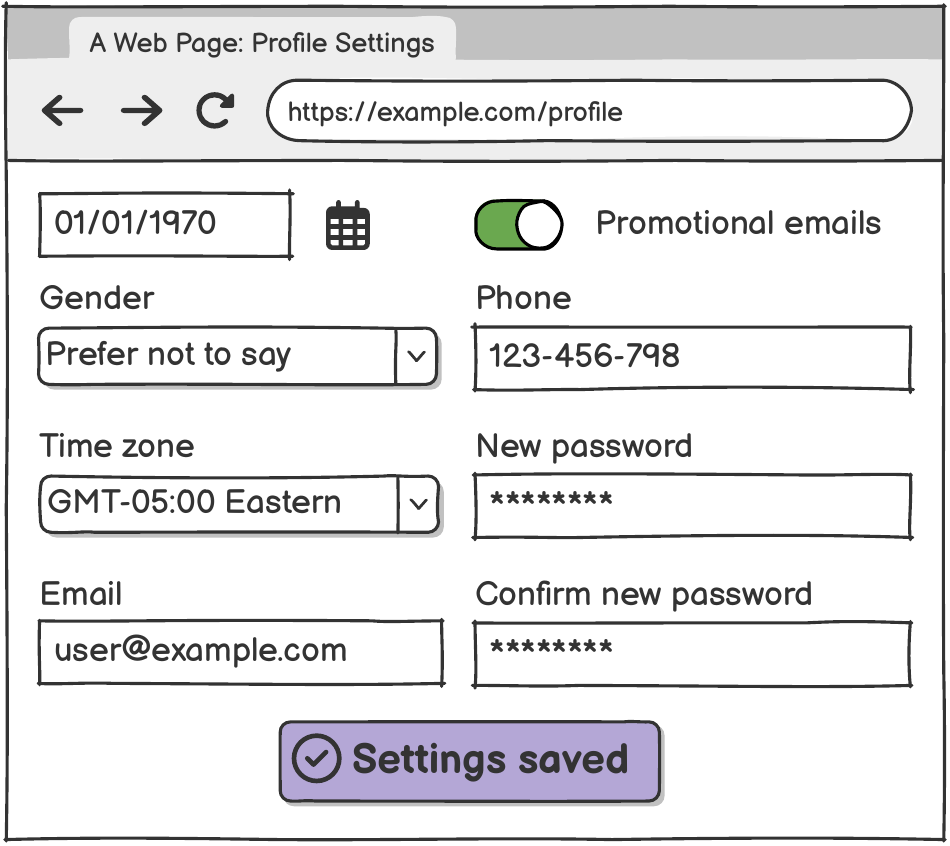}
         \caption{Button (ID25)}
         \label{fig:notification3}
     \end{subfigure}
          \begin{subfigure}[b]{0.50\columnwidth}
         \centering
         \includegraphics[width=\columnwidth]{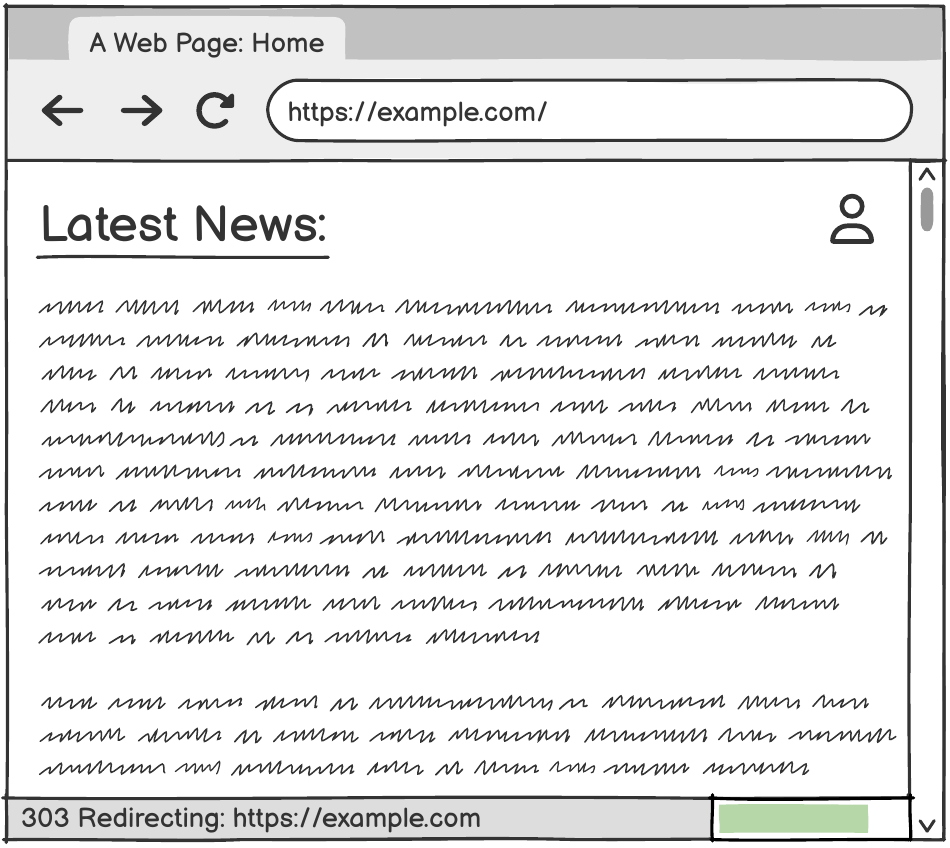}
         \caption{Redirect (ID28)}
         \label{fig:notification4}
     \end{subfigure}
        \caption{Common visual \textbf{feedback} we observed in \gls{pwc} and \gls{pwr} processes. (a) Illustrates an \textbf{explicit} password change \textbf{message}, (b) shows a tiny, quickly \textbf{disappearing cue} about ``Changes'' that have been saved, (c) \textbf{abuses a button} to inform the user about saved ``Settings,'' (d) does not display any text and instead \textbf{redirects} the user to the start page.}
        \label{fig:notifications}
\end{figure*}

\noindent \textbf{Password Reuse Prevention (ID17).}
About half of the websites (\gls{pwc}: 48, 50\%; \gls{pwr}: 47, 43\%; ID17) tried to mitigate the risk caused by reusing an old password.
For this, most websites use the ``current-password'' field in the \gls{pwc} process to compare the current and the new password.
We also found websites that restricted reuse of the current password, for example, by requiring users to change more than five characters.
Performing such similarity checks on cryptographically hashed passwords is challenging to implement and raises concerns about their secure storage.
Such checks can also impact user experience by interfering with users' password choices, potentially causing additional unnecessary mental load and memorability issues.\\

\noindent \textbf{Additional Security Advice (ID15).}
Some websites (\gls{pwc}: 20, 21\%; \gls{pwr}: 19, 17\%; ID15) provide advice while updating. 
This includes general security advice to protect user accounts, e.g., not to share passwords with others or how to create secure passwords, such as specific password lengths or using special characters (see Figure~\ref{fig:advice-etsy} in Appendix~\ref{app:website_examples}).

\subsection{Feedback After Setting a New Password}
\label{sec:visual-feedback}
We observed that websites implemented different types of feedback.
Below, we describe our findings on visual feedback, browser redirects, and notifications sent via email or SMS.
Some websites combined methods.
We visualize some representative examples in Figure~\ref{fig:notifications}.\\

\noindent \textbf{Visual Feedback and Redirects (ID22--30).}
Browser feedback included messages confirming a successful update, such as flash messages or redirects to a new page. Most sites displayed at least some kind of visual feedback after an update (\gls{pwc}: 78, 81\%; \gls{pwr}: 74, 68\%; ID22).
Many websites displayed feedback and then redirected to another page, while others did the opposite. 
A few (\gls{pwc}: 15, 16\%; ID29) redirected to the same page without displaying visual feedback; others did not provide any. 
Many times, we filled out the form and confirmed the update by clicking the \textit{submit} button without receiving any feedback.
Hence, it was unclear whether the update was successful.
The place and duration of flash messages varied widely.
In the \gls{pwc} process, many websites, and in the \gls{pwr} process, about half of the websites implemented messages as a persistent part of the website, directly above or below the password update form (\gls{pwc}: 35, 36\%; \gls{pwr}: 59, 54\%; ID25).
Other messages were hard to catch, as they quickly faded.
Most websites showed temporary messages for around one to two seconds (\gls{pwc}: 18, 19\%; \gls{pwr}: 8, 7\%; ID24).
Some required users to actively acknowledge and close them.\\

\noindent \textbf{Notifications via Email and SMS (ID31--32).}
During the update process, websites either sent an email or SMS notification or did not notify us at all.
Some notified us of a login (\gls{pwc}: 17, 18\%; \gls{pwr}: 6, 6\%; ID31), while the majority notified us of a successful password update (\gls{pwc}: 62, 65\%; \gls{pwr}: 64, 59\%; ID32).
Sporadically, we received notifications about a login from a new device or location.

\subsection{Automated Password Update Challenges}
\label{sec:automation_difficulties}
We now report findings that affect password managers' ability to automate password updates.
We found that many websites do not adhere to modern web standards and best practices for their password update forms~\cite{huaman-21-technical-issues}, making automated password updates unnecessarily challenging.\\

\noindent \textbf{Adoption of the ``Password-Change'' W3C Draft.}
While we analyzed all websites in our sample regarding their use of the W3C draft for changing passwords~\cite{mondello-24-well-known} with the ``.well-known'' standard~\cite{nottingham-19-rfc8615}, we found that only~\var{wellknowntotal} of \var{total_URLS} (5\%) implemented it. 
\var{wellknownchange} websites redirected to the \gls{pwc} process.
Another \var{wellknownrecovery} websites redirected to the \gls{pwr} process.
From those that redirected us to the \gls{pwc} process, only one website redirected us to the form where we could change the password. 
All the other websites redirected us to an account security page, where we had to navigate further to the actual password update form.\\

\noindent \textbf{HTML Attributes (ID4,9).}
Upon further investigation, we discovered that websites lacked support for autocomplete HTML attributes.
Only \var{html_autocomplete_both_True_df_change} of \var{total_URLS} (8\%) implemented the autocomplete attribute correctly in all login and password update form fields. 
The benefit of automation on these websites is huge since password managers can automatically fill every necessary form field during the \gls{pwc} process.
We also found that some websites supported only specific processes, such as the login process (43, 39\%; ID4), or only the update form during the \gls{pwc} process (\gls{pwc}: 15; ID9).

Furthermore, we observed that the autocomplete flow was incomplete on almost all websites (102 of \var{total_URLS}, 92\%), e.g., the autocomplete attribute was only available on the username or was missing in one of the fields in the change form. 
Moreover, some websites used non-standardized website-specific names or identifiers for relevant forms, making it difficult to automatically distinguish between the current and new password fields, e.g., ``ooui-php-1'' for a new password field or ``from-textbox 16899'' for the current password.
Additionally, we noticed that some websites had misleading HTML names, such as ``text'' for the username.
Some websites posed challenges due to missing field identifiers (names, IDs, autocomplete attributes). We even observed dynamically changing identifiers between page loads, likely from A/B testing, which occasionally caused autocomplete attribute conflicts, such as the unintended mapping of current password fields to new password inputs.
A few websites implemented the HTML ``passwordrules'' attribute~\cite{apple-20-manager-quirks}.
This attribute can support password managers in generating policy-compliant random passwords.

\section{Discussion}\label{sec:discussion}
Next, we discuss our findings and their implications for password update automation, end-user usability, developers, and web standardization, and contextualize them with prior work.
Our study descriptively measures deployed update workflows; without a user study, usability statements should be read as hypotheses grounded in established principles and prior work.
Where possible, we link observed design patterns to potential consequences, such as update errors, reduced password manager effectiveness, and weaker account security.

\subsection{Heterogeneity and Security are Obstacles} %
\label{sec:discussion-no-shortcuts}
Fully automating password updates could substantially streamline the process.
Our work illustrates many challenges for automating updates: 
Most sites did not implement the ``change-password'' W3C draft (6 of 111, 5\%) and made limited use of machine-readable interface practices (ID4, ID9).
Both are optional proposals (a working draft and vendor conventions), so we report adoption descriptively: non-adoption is not a deficiency per~se, but limits automation.
Furthermore, no website offered an API to update passwords.\\

\noindent \textbf{Interfaces Target Humans.}
Our analysis showed that current processes target humans.
They largely did not implement common practices for machine-readable interfaces~\cite{seeman-18-gap-analysis, chromium-16-password-form}, e.g., HTML attributes. 
Only \var{html_autocomplete_both_True_df_change} websites implemented attributes to support password managers in the login and password update forms.
A plausible cause is that machine-readability is not a design goal: flows are built and tested for humans, and tool support yields little direct benefit.
Dashlane's discontinued password changer~\cite{peters-21-dashlane-pwchanger-upgrade,dashlane-22-pwchanger-discontinued,sumner-23-dashlane-reasons} illustrates that automation requiring constant maintenance of site-specific heuristics is not sustainable at scale.
Our findings show that without adoption of optional aids such as the ``change password'' W3C draft, even automatically navigating to the update form is challenging~\cite{oh-25-checkup} (see Section~\ref{sec:automation_difficulties}).
Recent industry efforts to automate credential updates~\cite{waichulis-26-apple-passwords-automatic-pw-change} indicate that updating a compromised password remains a relevant target for automation.
Our results suggest that broader deployment of such functionality would benefit from machine-readable account-management interfaces rather than continued reliance on reverse-engineering human-facing workflows.
Currently, the names of links and paths to the update form vary significantly, making automated navigation difficult.
Forms often do not include HTML attributes such as ``autocomplete'' or ``passwordrules.''
Many sites also do not assign stable, descriptive IDs and names.
The user-facing consequence: when a password manager cannot fill or generate the new password, users may fall back to manually created passwords, which can be weaker or reused, or abandon the update~\cite{huaman-21-technical-issues, oesch-20-pw-managers}.
Websites also confuse password managers by failing to communicate policies without prior (failed) user interaction (see Section~\ref{sec:ressults_supportive_measures}).
Thus, we suggest that developers carefully test their forms with password managers to verify their implementation~\cite{huaman-21-technical-issues}.\\

\noindent \textbf{Gatekeepers Block Automation.}
Many of the tested websites deployed RBA, 2FA, or CAPTCHAs (see Section~\ref{sec:ressults_supportive_measures}).
While these measures aim to enhance security, they require user interaction and hinder automated processing.
This restricts password managers, but it reflects a genuine trade-off rather than a flaw: the same mechanisms block credential stuffing. Any future interface (see Section~\ref{sec:discussion-recommentadtion}) must preserve these protections; likewise, automation-friendly design must not come at the expense of users who create passwords manually.

\subsection{Inconsistent Patterns Overwhelm Users}
\label{sec:discussion-change-quest} 
Although all analyzed update processes targeted end-users, we identified multiple patterns that, following established usability heuristics, may impair usability and security. Our dataset contained no two sites with the same update pattern; instead, sites differed in navigation paths, update forms, and security measures.
The lack of common patterns across sites limits end-users' ability to transfer their experience and knowledge from one site to another.
Our findings suggest that each update process is a new quest, potentially involving complex challenges~\cite{bhattacharya-26-asi}.
Although end-users might be able to develop heuristics such as looking for terms like ``account,'' or ``security settings,'' the inconsistencies in the deployed processes cause unnecessary difficulties for end-users.
Concretely, missing confirmation fields or visibility toggles (ID8, ID11) can lead to unintended new passwords and lost account access; sessions on other devices usually remain valid after an update (terminated on only 8 of 96 PWC and 7 of 109 PWR websites; ID18), so a new password does not necessarily end unauthorized access; and missing feedback or notifications (ID22 to ID32) leave users unsure whether the update succeeded and let unauthorized changes go unnoticed.
(See the consequences outlined above and Section~\ref{sec:visual-feedback}.)\\

\noindent \textbf{Visibility Toggles.}
To protect against shoulder surfing attacks~\cite{aviv-17-shoulder-surfing-baseline}, most websites hide the entered password.
However, setting a new password can be error-prone, especially when using a software keyboard on a mobile device.
While our evaluation was limited to desktop browsers and mobile interfaces may differ, we found that some websites provided neither password confirmation fields nor visibility toggles.
Previous work suggests that offering a password visibility toggle encourages users to choose stronger passwords on mobile devices~\cite{melicher-16-mobile-passwords}.
To reduce end-user frustration, we recommend implementing a visibility toggle to improve the password update process.\\

\noindent \textbf{Strength Meters.}
Password strength meters that promote weak passwords might lead to a user's false sense of security~\cite{golla-18-psm}.
We noticed that certain strength meters did not behave correctly when dealing with copying and pasting or auto-filled passwords generated by a password manager, e.g., generated passwords may not be considered secure by a website's self-implemented password strength meter.\\

\noindent \textbf{Policies.}
Most websites deploy password policies to avoid weak passwords. 
Adhering to password policies can be challenging for both end-users and password managers. 
In particular, the wide variety of deployed password policies poses a significant challenge.
For password managers, generating a secure password according to the password policy of websites is not always possible, mostly due to certain length requirements or special character requirements~\cite{gautam-22-pw-policy-language}.
Moreover, these policies are almost never machine-readable due to the absence of the ``passwordrules'' HTML attribute~\cite{apple-18-password-rules}. 
Transparent, easy-to-understand, and early communication of the policy can reduce obstacles during updates, i.e., by ensuring the user sees and understands the policy when visiting the password update form without additional interactions~\cite{tan-20-pw-recommendations}.

\noindent \textbf{Generators.}
Some websites have embedded password generators and suggest them to users during the password update. 
While embedding generators into password update forms might help users choose stronger passwords or adhere to password policies, they often come with a usability drawback: %
Generated passwords are often difficult to memorize and manage~\cite{lyastani-18-pw-manager}, which makes good password manager support even more critical.

\noindent \textbf{Compromised Credential Checking (C3).}
We observed that few websites used password leak checks~\cite{hunt-18-pwned-passwords}. 
A password leak check can be a helpful feature to increase account security and protect users against using breached or compromised credentials, and is recommended for implementation by NIST~\cite{nist-25-sp800-63b-4}.
One limitation of the checks that prior work has observed is that they are usually strictly limited to the time when the password is updated~\cite{nisenoff-23-reuse}.\\

\noindent \textbf{Session Expiration.}
After a \gls{pwc} or \gls{pwr}, some websites require or enforce terminating all active sessions on all devices and browsers.
This feature is important, as it can drastically increase account security and is very welcome for ending online account sharing~\cite{obada-obieh-20-account-sharing}. 
Without it, an attacker might reuse an old active session after the account owner changes their password. 
User preferences may differ by update reason (e.g., routine change vs.\ unsharing), suggesting a user-controlled option (see R1).
Moreover, the efficacy of session termination as a security measure against attackers depends on the interplay with notification and confirmation practices for password and email address updates.\\

\noindent \textbf{Supporting Users.}
Websites utilize various tools to enhance account security, with many supporting updates through confirmation fields, visibility toggles, feedback, and post-change notifications.
However, critical gaps persist in \gls{pwc} and \gls{pwr} workflows: while most sites communicate a policy (ID10), strength meters appear on about a third (ID13), and generators (ID12) and leak checks (ID14) are almost absent (2 and 1 of 111).
While simplification of password updates is evident, robust assistance for creating secure credentials remains inconsistently implemented, highlighting a disconnect between usability efforts and proactive security reinforcement.

\subsection{Recommendations}
\label{sec:discussion-recommentadtion}
We organize our recommendations by priority, ground each in Table~\ref{tab:obstacles:sec_measures}, and then discuss the actors best positioned to act.\\

\noindent \textbf{High Priority:} Security-critical.

\vspace{0.1em}\recitem{\textbf{R1: Treat Updates as Revocation, Not Replacement.}\\
Only 8/96 (\gls{pwc}) and 7/109 (\gls{pwr}) sites terminated all sessions (ID18). Lingering sessions let attackers persist despite a new password. List and offer to expire other sessions, tokens, and cookies in every update flow.}

\vspace{0.1em}\recitem{\textbf{R2: Always Notify the User Out of Band.}\\
34/96 (\gls{pwc}) and 45/109 (\gls{pwr}) sites do not notify users about the update (ID32), enabling stealthy changes. Send helpful notifications with a remediation option.}

\vspace{0.1em}\recitem{\textbf{R3: Provide Guided Account Remediation.}\\
Guide users through related account-security steps, including active sessions, recovery, and 2FA, rather than treating the password update as a complete action.}

\clearpage
\noindent \textbf{Medium Priority:} Prevents errors and failed updates.

\vspace{0.1em}\recitem{\textbf{R4: Make Password Updates Unambiguous.}\\
Provide a confirmation field or visibility toggle, missing on 21 and 43 of 96 \gls{pwc} forms, respectively (ID8,11), visual feedback (ID24--26), and display the policy (ID10).}

\vspace{0.1em}\recitem{\textbf{R5: Support and Test Password Manager Autofill.}\\
Only 9/111 sites used autocomplete correctly (ID4,9).}

\vspace{0.5em}
\noindent \textbf{Lower Priority:} Optional ecosystem adoption.

\vspace{0.1em}\recitem{\textbf{R6: Implement the Change-Password URL.}\\
Only 6/111 sites adopted this proposal. This redirect can remove the navigation hurdle of two to six steps.}

\vspace{0.1em}\recitem{\textbf{R7: Evolve Standards Toward a Guarded API.}\\
An API must preserve 2FA, RBA, and anti-abuse controls.}
~

\noindent\textbf{Developers, Frameworks, Operators, and Vendors.}
Our results suggest that design choices in websites and frameworks significantly contribute to end-user usability and automation capabilities, and that both can be substantially improved.
Since developers struggle with usable security features and advice~\cite{gutfleisch-22-non-usable-software, klemmer-23-every-week, nguyen-17-stitch-in-time, acar-17-crypto-apis}, frameworks must ship secure and usable defaults.
Hence, we recommend that web framework developers better support usability and the interaction between password managers and password change and reset processes (R4, R5).
In particular, they should improve HTML attributes such as ``name'', ``id'', ``autocomplete'', ``pattern'', and even ``passwordrules'' to make them the new default, i.e., standardized form templates that follow best practices~\cite{django-26-auth-form, chromium-16-password-form, apple-18-password-rules}.
Frameworks should implement the ``change-password'' W3C draft (R6)~\cite{mondello-24-well-known}. In addition, server operators and web servers~\cite{apache-95-apache-webserver, sysoev-04-nginx-webserver} themselves could further lower the burden via dedicated directives, as improved tooling is known to increase adoption~\cite{tiefenau-19-lets-encrypt, kitamura-20-developer-well-known}.
Major browsers like Google~Chrome could drive automated password reset adoption, mirroring their successful HTTPS standardization efforts that labeled HTTP as ``not secure''~\cite{schechter-18-milestone-in-https}.\\

\noindent \textbf{An Extended Standard for Password Managers.}
While the ``change-password'' W3C draft aims to support websites, end-users, and password managers in deploying more usable password update processes, the standard has limitations and lacks widespread adoption.
The standard is limited to redirecting end-users to the website's actual password change URL.
Even where deployed, managers must still parse update forms that are often difficult to process automatically.
Overall, our findings suggest that requiring password managers to interact with password change and reset processes primarily designed for end-user interactions has limited chances of success.
Instead, we recommend extending the ``change-password'' W3C draft to a more powerful API that allows browsers and password managers to automatically discover a website's password change URL and complete the process (R7). 
An effective API must balance password policy compliance~\cite{dashlane-21-well-known-proposal}, server-side security controls, 2FA, and anti-abuse protections---a complex challenge requiring cryptographic, security, and HCI expertise.
In particular, it must not weaken the gatekeepers discussed in Section~\ref{sec:discussion-no-shortcuts} (user-initiated, rate-limited).
More work is needed to better support developers in implementing password manager-friendly processes.\\

\noindent \textbf{Recommendations for Future Research.}
Our findings revealed multiple patterns for \gls{pwc} and \gls{pwr} processes.
While the considerable heterogeneity of patterns negatively impacts users' abilities to transfer experiences and knowledge between websites, it is also unclear which \gls{pwc} and \gls{pwr} processes provide end-users the best usability and security.
As our study is descriptive, a user study should test the usability hypotheses derived from our measurements (e.g., whether deeper paths, missing confirmation fields, or ephemeral feedback increase errors and abandonment) and identify the most promising process designs.
Previous work~\cite{klivan-23-cred-management} has shown that end-users distrust password managers. 
Future research could focus on understanding and improving the acceptance of automated password changes, particularly in the context of browsers and password managers.
Future work could also examine the root causes of the limited adoption of such standards and how future standards can be improved to increase adoption.

\section{Conclusion}\label{sec:conclusion}

``\emph{So what makes changing a password hard?}''
Our analysis of 111~websites shows that the difficulty lies not only in the password itself, but in fragmented account-management processes.
In security-relevant situations, such as suspected compromise, users face hidden entry points, inconsistent labels, unclear policies, and ambiguous outcomes.
As a result, they may avoid password updates altogether.
Our findings suggest that password updates should be treated as security-critical workflows.
Websites should provide direct update paths, communicate requirements before users fail, clearly confirm successful changes, notify users out of band, and support expiring old sessions.
They should also expose predictable, machine-readable interfaces for password managers.
Yet, only a few websites implemented the well-known change-password URL or correctly supported password autofill across login and update forms.
Secure password remediation, therefore, requires ecosystem-level support from developers, web frameworks, and standards bodies, so that password updates become reliable and usable when users need them most.

\clearpage
\section*{Acknowledgments}

\label{bp:ack_reviewers}
We remember our co-author Marten Oltrogge with deep gratitude.
Marten made important contributions to the ideas, analysis, and writing of this paper.
Sadly, he passed away before the paper was submitted and finalized.
We dedicate this work to his memory; any remaining errors are our own.\\
~\\
This research was partly funded by VolkswagenStiftung Niedersächsisches Vorab (ZN3695) and the European Union as part of the program Digital Europe.
Neither the European Union nor the other granting authorities can be held responsible.
Any findings and opinions expressed in this material are those of the authors and do not necessarily reflect the views of the funding agencies.

\makeatletter
\interlinepenalty=10000
\bibliographystyle{plain}
\bibliography{references/bibliography}
\makeatother
\clearpage

\appendix

\clearpage
\onecolumn
\section{Analyzed Websites}
\label{appendix:full-websites-list}
\begin{table}[!ht]
\centering
\caption{Full list of analyzed websites, their Tranco rank, and website category (extracted from \href{https://alphamountain.ai}{alphamountain.ai}).}
\label{tab:analyzed-websites}
\scriptsize
\setlength{\tabcolsep}{2.5pt}
\renewcommand{\arraystretch}{0.92}
\resizebox{\textwidth}{!}{%
\rowcolors{2}{gray!10}{white}
\begin{tabular}{r l l@{\hspace{1.4em}}r l l@{\hspace{1.4em}}r l l}
\toprule
\textbf{Rank} & \textbf{Website} & \textbf{Category} & \textbf{Rank} & \textbf{Website} & \textbf{Category} & \textbf{Rank} & \textbf{Website} & \textbf{Category} \\
\midrule
    1 & google.com & Search Engines/Portals & 140 & quora.com & Reference & 283 & walmart.com & Shopping \\
    7 & twitter.com & Social Networking & 141 & stackoverflow.com & Information Technology & 286 & cnbc.com & Finance \\
    12 & apple.com & Information Technology & 145 & roblox.com & Games & 288 & hubspot.com & Information Technology \\
    13 & cloudflare.com & Information Technology & 147 & theguardian.com & News & 290 & unsplash.com & Media Sharing \\
    14 & linkedin.com & Social Networking & 150 & bbc.com & News & 298 & unity3d.com & Information Technology \\
    16 & wikipedia.org & Reference & 152 & spankbang.com & Pornography & 299 & mediafire.com & File Sharing/Storage \\
    18 & live.com & Search Engines/Portals & 159 & sciencedirect.com & Education & 302 & espn.com & Sports \\
    20 & amazon.com & Shopping & 162 & etsy.com & Shopping & 316 & onlyfans.com & Adult/Mature \\
    31 & github.com & Information Technology & 163 & twitch.tv & Video/Multimedia & 319 & deviantart.com & Arts/Culture \\
    32 & pinterest.com & Social Networking & 166 & sourceforge.net & Software Downloads & 327 & dailymotion.com & Video/Multimedia \\
    33 & reddit.com & Forums & 168 & imgur.com & Media Sharing & 330 & grammarly.com & Productivity Applications \\
    34 & wordpress.org & Information Technology & 171 & shopify.com & Information Technology & 332 & figma.com & Information Technology \\
    41 & fastly.net & Content Servers & 174 & hp.com & Information Technology & 333 & ft.com & Finance \\
    42 & zoom.us & Virtual Meetings & 181 & dell.com & Information Technology & 335 & fiverr.com & Business/Economy \\
    44 & mail.ru & Search Engines/Portals & 186 & oracle.com & Information Technology & 337 & w3schools.com & Education \\
    46 & adobe.com & Information Technology & 192 & epicgames.com & Games & 342 & eventbrite.com & Entertainment \\
    48 & yandex.ru & Search Engines/Portals & 199 & issuu.com & Productivity Applications & 347 & snapchat.com & Social Networking \\
    54 & gandi.net & Hosting & 200 & tradingview.com & Brokerage/Trading & 351 & npr.org & News \\
    56 & intuit.com & Finance & 211 & deepl.com & Translation & 358 & time.com & News \\
    66 & tiktok.com & Social Networking & 216 & discord.com & Chat/IM/SMS & 361 & surveymonkey.com & Business/Economy \\
    78 & pornhub.com & Pornography & 218 & pixiv.net & Arts/Culture & 364 & chaturbate.com & Pornography \\
    82 & paypal.com & Finance & 222 & alibaba.com & Shopping & 365 & pixabay.com & Media Sharing \\
    83 & tumblr.com & Social Networking & 223 & washingtonpost.com & News & 368 & telegraph.co.uk & News \\
    96 & nytimes.com & News & 226 & reuters.com & News & 371 & goodreads.com & Reference \\
    97 & fandom.com & Entertainment & 229 & ibm.com & Information Technology & 372 & scribd.com & Reference \\
    99 & dropbox.com & File Sharing/Storage & 231 & samsung.com & Business/Economy & 380 & notion.so & Information Technology \\
    105 & ebay.com & Auctions/Classifieds & 238 & bloomberg.com & Finance & 381 & yelp.com & Reference \\
    106 & webex.com & Virtual Meetings & 245 & wix.com & Hosting & 386 & no-ip.com & Dynamic DNS \\
    108 & imdb.com & Entertainment & 248 & atlassian.com & Information Technology & 387 & sentry.io & Information Technology \\
    109 & flickr.com & Media Sharing & 250 & linktr.ee & Information Technology & 389 & name.com & Hosting \\
    113 & naver.com & Search Engines/Portals & 252 & reg.ru & Hosting & 392 & ted.com & Education \\
    120 & soundcloud.com & Audio & 267 & stripe.com & Finance & 395 & themeforest.net & Information Technology \\
    121 & digicert.com & Information Technology & 275 & nature.com & Education & 396 & statista.com & Reference \\
    122 & aliexpress.com & Shopping & 276 & ilovepdf.com & Productivity Applications & 397 & wired.com & News \\
    127 & xhamster.com & Pornography & 278 & kaspersky.com & Information Technology & 400 & typeform.com & Productivity Applications \\
    129 & archive.org & Reference & 281 & businessinsider.com & News & 401 & binance.com & Finance \\
    137 & meraki.com & Information Technology & 282 & cloudns.net & Hosting & 402 & rakuten.co.jp & Shopping \\
\bottomrule
\end{tabular}
\rowcolors{2}{}{}%
}
\end{table}
\normalsize

\vspace{0.5em}

\captionsetup{hypcap=false}

\setlength{\columnsep}{1.5em}

\begin{paracol}{2}

\section{Used Quantifiers}\label{app:quantifiers}

\begin{center}
\resizebox{\linewidth}{!}{%
    \begin{tikzpicture}
    \def\websites{111} %
    \def\spacing{2} %
    \def\numLines{7} %
    \def\vertLength{1} %

    \pgfmathsetmacro{\none}{int(0)}
    \pgfmathsetmacro{\afew}{int(floor(\websites * 0.15))}
    \pgfmathsetmacro{\some}{int(floor(\websites * 0.30))}
    \pgfmathsetmacro{\many}{int(floor(\websites * 0.45))}
    \pgfmathsetmacro{\abouthalf}{int(floor(\websites * 0.55))}
    \pgfmathsetmacro{\majority}{int(floor(\websites * 0.70))}
    \pgfmathsetmacro{\most}{int(floor(\websites * 0.85))}
    \pgfmathsetmacro{\almostall}{int(floor(\websites * 0.99))}
    \pgfmathsetmacro{\all}{int(\websites)}
    \pgfmathsetmacro{\noneplus}{int(\none + 1)}
    \pgfmathsetmacro{\afewplus}{int(\afew + 1)}
    \pgfmathsetmacro{\someplus}{int(\some + 1)}
    \pgfmathsetmacro{\manyplus}{int(\many + 1)}
    \pgfmathsetmacro{\abouthalfplus}{int(\abouthalf + 1)}
    \pgfmathsetmacro{\majorityplus}{int(\majority + 1)}
    \pgfmathsetmacro{\mostplus}{int(\most + 1)}
    \pgfmathsetmacro{\almostallplus}{int(\almostall + 1)}

    \draw[thick] (0, 0) -- (\numLines*\spacing, 0);

    \foreach \i in {0,...,\numLines} {
        \ifnum\i=0
            \draw[thick] (\i*\spacing, -\vertLength) -- (\i*\spacing, \vertLength); %
        \else\ifnum\i=\numLines
            \draw[thick] (\i*\spacing, -\vertLength) -- (\i*\spacing, \vertLength); %
        \else
            \draw[thick, dotted] (\i*\spacing, -\vertLength) -- (\i*\spacing, \vertLength); %
        \fi\fi

    }

    \node[below] at (0*\spacing, -\vertLength) {0\%};
    \node[below] at (1*\spacing, -\vertLength) {15\%};
    \node[below] at (2*\spacing, -\vertLength) {30\%};
    \node[below] at (3*\spacing, -\vertLength) {45\%};
    \node[below] at (4*\spacing, -\vertLength) {55\%};
    \node[below] at (5*\spacing, -\vertLength) {70\%};
    \node[below] at (6*\spacing, -\vertLength) {85\%};
    \node[below] at (7*\spacing, -\vertLength) {100\%};

    \node at (-0.3*\spacing, -0.5*\vertLength) {None};
    \node at (0.5*\spacing, -0.5\vertLength) {A few};
    \node at (1.5*\spacing, -0.5*\vertLength) {Some};
    \node at (2.5*\spacing, -0.5*\vertLength) {Many};
    \node at (3.5*\spacing, -0.5*\vertLength) {About Half};
    \node at (4.5*\spacing, -0.5*\vertLength) {Majority};
    \node at (5.5*\spacing, -0.5*\vertLength) {Most};
    \node at (6.5*\spacing, -0.5*\vertLength) {Almost All};
    \node at (7.3*\spacing, -0.5*\vertLength) {All};

\end{tikzpicture}

}
\captionof{figure}{Quantifiers used throughout our results.}
\label{fig:quantifier-figure}
\end{center}

\vspace{0.75em}

\section{Example Screenshots}
\label{app:website_examples}

\begin{center}
\includegraphics[width=0.75\linewidth]{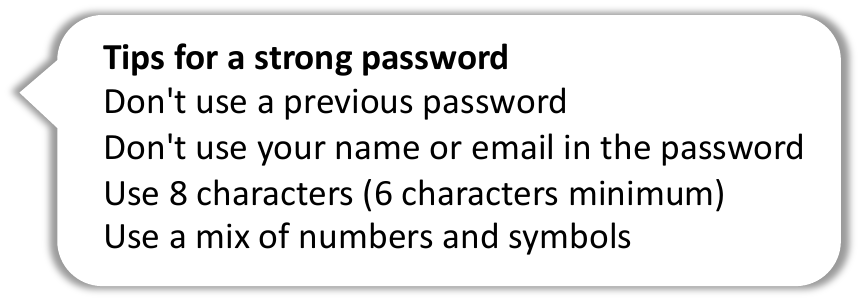}
\captionof{figure}{Advice when changing the password on \href{https://etsy.com}{etsy.com}.}
\label{fig:advice-etsy}
\end{center}

\switchcolumn

\vspace{4em}

\begin{center}
\includegraphics[width=0.95\linewidth]{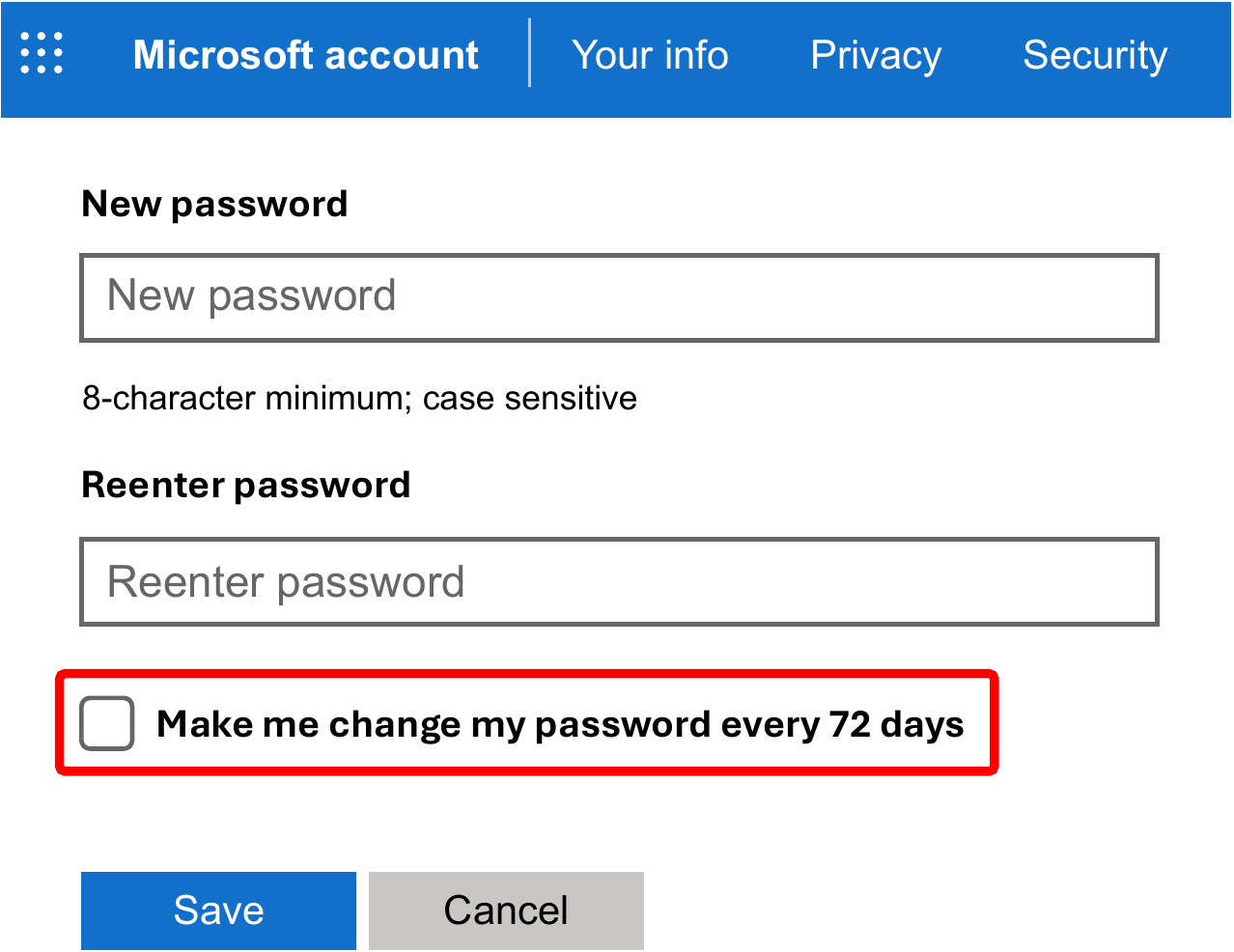}
\captionof{figure}{Optional expiration setting on \href{https://microsoft.com}{microsoft.com}.}
\label{fig:expiration}
\end{center}

\end{paracol}

\captionsetup{hypcap=true}

\end{document}